\documentclass[%
 reprint,
 amsmath,amssymb,
pre
]{revtex4-2}

\usepackage{graphicx}
\usepackage{dcolumn}
\usepackage{bm}

\usepackage{amsfonts,xcolor,soul}
\usepackage{ulem}
\usepackage{float}
\usepackage{hyperref}
\usepackage{subfigure}
\usepackage{comment}

\newcommand{\p} {\partial}
\newcommand{\el}{\boldsymbol{\ell}}
\newcommand{\nab}{\boldsymbol\nabla}
\newcommand{\elb}{\boldsymbol\ell}
\def\vv{{\bf v}}
\def\vvprim{{\bf v'}}

\def\bb{{\bf v_A}}
\def\bbprim{{\bf v'_A}}
\def\rr{{\bf \el}}
\def\div{{\pmbmath{\nabla} \cdot}}
\def\divprim{{\pmbmath{\nabla}' \cdot}}
\def\divl{{\pmbmath{\nabla}_\rr \cdot}}
\def\grad{{\pmbmath{\nabla}}}
\def\gradprim{{\pmbmath{\nabla}'}}

\def\jj{\mathbf{J}}
\def\jjc{\mathbf{J_c}}
\def\jjcprim{\mathbf{J'_c}}

\def\Bdir{\mathbf{b}}

\def\pmat{\overline{\overline{P}}}
\def\pmatprim{\overline{\overline{P'}}}
\def\ptot{\overline{\overline{P_*}}}
\def\ptotprim{\overline{\overline{P'_*}}}
\def\id{\overline{\overline{I}}}
\def\pperp{P_{\perp}}

\def\ppar{P_{\parallel}}

\def\({\left(}
\def\){\right)}
\def\langle{\left<}
\def\rangle{\right>}
\def\[{\left[}
\def\]{\right]}

\def\be{\begin{equation}}
\def\ee{\end{equation}}
\def\ba{\begin{eqnarray}}
\def\ea{\end{eqnarray}}

\def \pmbmath{\mathpalette\pmbmathaux}
\def \pmbmathaux#1#2{
         \pmbtext{$#1#2$}}
\def \pmbtext#1{\leavevmode
     \setbox0\hbox{#1}
     \kern0,4pt \copy0 \kern-\wd0
     \kern-0,2pt \raise0,3pt \box0 }

\begin{document}

\preprint{APS/123-QED}

\title{Exact law for compressible pressure-anisotropic magnetohydrodynamic turbulence: toward linking energy cascade and instabilities}

\author{P. Simon} 
\email{pauline.simon@lpp.polytechnique.fr}
\author{F. Sahraoui} 
\affiliation{Laboratoire de Physique des Plasmas, CNRS, \'Ecole polytechnique, Universit\'e Paris-Saclay, Sorbonne Universit\'e, Observatoire de Paris-Meudon, F-91128 Palaiseau Cedex, France}

\date{\today}
\begin{abstract}
We derive a first exact law for compressible pressure-anisotropic magnetohydrodynamic turbulence. For a gyrotropic pressure tensor, we study the double-adiabatic case and show the presence of new flux and source terms in the exact law, reminiscent of the plasma instability conditions due to pressure anisotropy. The Hall term is shown to bring ion-scale corrections to the exact law without affecting explicitly the pressure terms. In the pressure isotropy limit we recover all known results obtained for isothermal and polytropic closures. The incompressible limit of the gyrotropic system leads to a generalization of the Politano and Pouquet's law where a new {\it incompressible} source term is revealed and reflects exchanges of the magnetic and kinetic energies with the no-longer-conserved internal energy.  We highlight the possibilities offered by the new laws to investigate potential links between turbulence cascade and instabilities widely observed in laboratory and astrophysical plasmas.  
\end{abstract} 


\maketitle

\section{Introduction}\label{sec:intro} 
In recent years there has been a growing interest in deriving Von-Karman-Howarth-Monin (vKHM) \citep{von_karman_statistical_1938,monin_statistical_1971,monin_statistical_1975} equations that describe turbulent energy cascade in magnetized plasmas. Those equations present the double advantage of being fully nonlinear and of linking the turbulent energy cascade (or dissipation) rate to measurable fields \citep{kolmogorov_dissipation_1991,kolmogorov_local_1991,frisch_turbulence_1995,antonia_analogy_1997,zakharov_kolmogorov_1992,galtier_introduction_2016,galtier_physique_2021}.The cascade rate, is used to estimate energy dissipation from spacecraft data taken in the solar wind (SW) and the planetary plasma environments \citep{smith_dependence_2006,sorriso-valvo_observation_2007,andres_energy_2019,sorriso-valvo_turbulence-driven_2019,bandyopadhyay_situ_2020,quijia_comparing_2021}. Efforts were thus put in generalizing the laws to more realistic conditions met in those plasmas at the cost of increasing complexity. Two main lines of research are pursued: one aiming at extending the range of the described scales, from magnetohydrodynamics (MHD), to Hall-MHD and two-fluids \citep{politano_dynamical_1998,politano_von_1998,galtier_von_2008,andres_exact_2016,andres_von_2016,banerjee_alternative_2017,hellinger_von_2018,ferrand_exact_2019}; the second by incorporating density fluctuations described within isothermal or polytropic closures \citep{galtier_exact_2011,banerjee_exact_2013,banerjee_kolmogorov-like_2014,banerjee_scaling_2016,andres_alternative_2017,andres_exact_2018,andres_energy_2018,banerjee_scale--scale_2020,ferrand_compressible_2020,simon_general_2021} or gravitational effects to study star formation in the interstellar medium \citep{banerjee_exact_2017,banerjee_energy_2018}.

Despite these important improvements, a key missing ingredient that none of the existing models can describe, is the presence of pressure anisotropy (with respect to the background magnetic field ${\bf B}_0$). Indeed, while the existing laws do consider the presence of a background magnetic field, which allows one to study energy transfers along the parallel and perpendicular directions to ${\bf B}_0$ \citep{macbride_turbulent_2008,stawarz_turbulent_2009,osman_anisotropic_2011,hadid_energy_2017,hadid_compressible_2018}, they however all assume a scalar pressure, which is unrealistic to describe most of magnetized collisionless astrophysical (or laboratory) plasmas where ion and electron pressure anisotropies are frequently reported from particle measurements \citep{griffel_anisotropy_1969,gary_proton_2001,kasper_windswe_2002,bale_magnetic_2009,hellinger_solar_2006, sahraoui_anisotropic_2006}.

In order to include pressure anisotropy in fluid modeling of magnetized plasmas, \citet{chew_boltzmann_1956} introduced the double-adiabatic closure (known also as CGL). One of the main changes to the dynamics of the plasma brought up by pressure anisotropy in CGL-MHD equations is the presence of instabilities, which in the linear limit coincide with the firehose when $\frac{\beta_{\parallel}}{2}[1-a_p] > 1$ and the mirror when $\beta_{\parallel} a_p + 1 < \frac{\beta_{\parallel} a_p^2}{6}$ ($\beta_{\parallel}$ is the ratio of the parallel thermal to the magnetic pressure, $a_p=T_\perp/T_\parallel$ is the ratio between the proton perpendicular and parallel temperatures \citep{passot_collisionless_2007,hunana_introductory_2019,hau_mhd_2007}). These instabilities (or their kinetic counterparts) were shown to constrain part of the dynamics of the SW \citep{hellinger_solar_2006,bale_magnetic_2009} and are thought to operate in laboratory devices \citep{scime_ion_2000}, clusters of galaxies \citep{schekochihin_magnetised_2005} and black holes' accretion disks \citep{sharma_shearing_2006}. However, the interplay between turbulence and instabilities remains an unsettled question although some hints were already reported. These include driving of sub-ion scale turbulence \citep{sahraoui_magnetic_2004,sahraoui_anisotropic_2006,kunz_firehose_2014}, influencing the scaling of the high frequency magnetic energy spectra in the SW \citep{bale_magnetic_2009}, or linking unstable plasmas to high energy cascade rates as measured in the near-Earth space \citep{osman_proton_2013, hadid_compressible_2018}, which remains to date not fully understood.

It is the goal of this work to fill the existing gap by providing a self-consistent (fluid) theoretical framework to investigate the potential coupling between plasma turbulence and instabilities.

\section{Theoretical model}
We use the classical MHD equations but assume a (symmetric) pressure tensor rather than a scalar one,
\ba
\label{eq:sys1} &\p_t& \rho = - \div (\rho \vv)  \, , \\
\label{eq:sys2} &\p_t& (\rho\vv) = \div (\rho \bb \bb-\rho \vv \vv-\ptot) + {\bf d_k} + {\bf f} \, , \\
\label{eq:sys4}&\p_t& (\rho \bb)  = \div (\rho \bb \vv - \rho \vv \bb) + \rho \vv \div \bb  \nonumber \\ 
&&- \frac{1}{2} \rho \bb \div \vv + {\bf d_m}  \, ,
\ea 
where $\rho$ is the mass density, $\vv$ the velocity field, $\bb = \frac{\mathbf{B}}{\sqrt{\mu_0\rho}}$ the Alfv\'en velocity with $\mathbf{B}$ the magnetic field, and $\ptot = \pmat + \pmat_M$ is the total pressure tensor, i.e., the sum of the pressure tensor $\pmat$  and the magnetic pressure tensor $\pmat_M = P_M \id = (\rho {v_A}^2/2) \id$ ($\id$ is the identity 3x3 matrix), ${\bf d_k}$ the kinetic viscous dissipation, ${\bf d_m}$ the magnetic diffusivity, ${\bf f}$ a stationary homogeneous external force assumed to act on the largest scales.

Since we want to derive the exact law for the total energy of the system, equations (\ref{eq:sys1})-(\ref{eq:sys4}) are complemented by that of the (specific) internal energy $u$, which reads
\be
\label{eq:sys3} \p_t u = - \div (u \vv ) + u \div \vv - \frac{\pmat}{\rho} : \grad \vv    \, ,
\ee 
since $\pmat$ and $\ptot$ are symmetrical tensors, i.e., $P_{ij}=P_{ji}$, the dual product between two of such tensors $\pmat$ and $\overline{\overline{A}}$ obeys $\pmat:\overline{\overline{A}}=P_{ij}A_{ij}=P_{ji}A_{ij}$. Equation (\ref{eq:sys3}), valid for any symmetric pressure tensor when the heat flux is neglected, can be derived from thermodynamical considerations \citep{hazeltine_local_2013} or from the moments of the Vlasov-Maxwell equations \citep{hunana_introductory_2019}. For a scalar pressure, i.e., $\pmat = P \id$, we recover the equation of the internal energy used in \cite{simon_general_2021}. 

Equations (\ref{eq:sys1})-(\ref{eq:sys3}) will be used in the following section to derive the exact law of interest.

\section{General  exact law for compressible pressure-anisotropic MHD turbulence}
Following the standard approach used in statistical theories of fully developed turbulence \citep{kolmogorov_local_1991,kolmogorov_dissipation_1991,frisch_turbulence_1995,galtier_introduction_2016}, we define the spatial increment (or scale) $\el$ connecting two points $\textbf{x}$ and $\textbf{x}'$ as ${\bf x'} = {\bf x}+{\bf \el}$ and introduce the notations, $\xi(\textbf{x}) \equiv \xi$, its conjugate (i.e., taken at the position $\textbf{x}'$) $\xi(\textbf{x}') \equiv \xi'$ and the incremental quantity $\delta \xi \equiv \xi'-\xi$. These definitions impose that $\partial_x \xi' = \partial_{x'} \xi = 0$, while the hypothesis of space homogeneity implies the relations $\langle\nab'\cdot\rangle=\nab_{\elb} \cdot\langle\rangle$ and $\langle\nab\cdot\rangle=-\nab_{\elb} \cdot\langle\rangle$, where $\nab_{\elb}$ denotes the derivative operator along the increment vector $\elb$ and $\langle \rangle$ an ensemble average.

We consider the mean correlation function of the total energy $R_{tot} =  (R + R') /2$ with $ R  = \langle \rho \vv \cdot \vv' /2 + \rho \bb \cdot \bb' /2 +  \rho u' \rangle =  R_k + R_B +  R_u $ a correlation function taken at the point $\textbf{x}$ and $R'$ its conjugate. We remark that if $\textbf{x} = \textbf{x}'$, $R_{tot} =  E  = \langle \rho v^2 /2 + \rho v_A{}^2 /2 + \rho u \rangle $, i.e. the mean total energy of the system. 

Using the property $\partial_t \langle \rangle = \langle \partial_t \rangle$, the equations (\ref{eq:sys1})-(\ref{eq:sys3}) written at the independent positions $\textbf{x}$ then $\textbf{x}'$ and multiplied by the appropriate variables (e.g., equation (\ref{eq:sys2}) multiplied by $\vv'$) and the space homogeneity assumption (see \cite{simon_general_2021} for more details), we obtain the temporal evolution of the kinetic, $ R_k $, the magnetic, $ R_B $, and the internal energy, $ R_u $, correlators:
\ba
&2&\p_t  R_k   = - \divl \langle \rho  \vv \cdot \vvprim \delta \vv +  \rho \bb \cdot \vvprim \bb - \rho \vv \cdot \bbprim \bbprim\rangle \nonumber\\ 
&& + \divl  \langle \ptot \cdot \vvprim  - \frac{\rho}{\rho'} \ptotprim \cdot \vv  \rangle+ \langle \rho \vv \cdot \vvprim \divprim \vvprim \rangle \nonumber\\ 
&&  - \langle \frac{\rho}{\rho'}\vv \cdot \ptotprim \cdot \frac{\gradprim \rho' }{\rho'} + 2  \rho \vv \cdot \bbprim \divprim \bbprim  \rangle + \mathcal{F} + \mathcal{D}_k \, , \label{eq:Rk} \\
&2& \p_t R_B  = -\divl \langle \rho  \bb \cdot \bbprim \delta \vv  + \rho \vv \cdot \bbprim\bb \rangle \nonumber\\ 
&& - \divl \langle \rho \bb \cdot\vvprim \bbprim \rangle + \langle  \frac{1}{2}\rho\bb \cdot \bbprim (\divprim \vvprim -\div \vv ) \rangle\nonumber\\ 
&& + \langle \rho \vv\cdot \bbprim \div \bb - \rho\bb \cdot \vvprim \divprim \bbprim \rangle  + \mathcal{D}_m\, ,  \label{eq:Rb} \\
&\p_t& R_{u} = - \divl \langle \rho u' \delta \vv \rangle + \langle \rho u' \divprim \vvprim - \frac{\rho}{\rho'} \pmatprim : \gradprim \vvprim   \rangle  \, , \label{eq:Ru}
\ea 
where the terms depending on the forcing, the kinetic, and magnetic dissipation are regrouped respectively in $\mathcal{F}$, $\mathcal{D}_k$ and $\mathcal{D}_m$. Then the temporal evolution of $R_{tot}$ is the sum of the relations (\ref{eq:Rk})-(\ref{eq:Ru}) and of their conjugates (written at position $\textbf{x}'$). By recognizing the developed form of the structure functions $\langle\delta \(\rho \vv\) \cdot \delta \vv \delta \vv \rangle $, $\langle\delta \(\rho \bb\) \cdot \delta \bb \delta \vv \rangle $, $\langle\delta \(\rho \bb\) \cdot \delta \vv \delta \bb \rangle $, $ \langle\delta \(\rho \vv\) \cdot \delta \bb \delta \bb \rangle$, $ \langle \delta \rho \delta u \delta \vv \rangle $, $\langle \delta \rho \delta (\pmat/\rho) \cdot \delta \vv \rangle$ and $\langle \delta \rho \delta (\pmat_M /\rho) \cdot \delta \vv \rangle$, the final expression for the temporal evolution of the total energy correlator reads:
\begin{widetext}
\ba
&4& \p_t R_{tot} = \divl \langle \(\delta \(\rho \vv\) \cdot \delta \vv  + \delta \(\rho \bb\) \cdot \delta \bb + 2 \delta \rho \delta u\) \delta \vv -\( \delta \(\rho \bb\) \cdot \delta \vv + \delta \(\rho \vv\) \cdot \delta \bb \) \delta \bb - \delta \rho \delta \(\frac{\ptot}{\rho}\)\cdot \delta \vv  \rangle\nonumber \\
&&\begin{split}
&+ \langle \(\rho \vv \cdot \delta \vv  +\frac{1}{2} \rho \bb \cdot \delta \bb - \frac{1}{2}\bb \cdot \delta \(\rho \bb\) +2 \rho \delta u\)\divprim \vvprim - 2\rho \delta \(\frac{\pmat}{\rho}\) : \gradprim \vvprim \rangle  \\
&+ \langle \( - \rho' \vvprim \cdot \delta \vv - \frac{1}{2} \rho'\bbprim \cdot \delta\bb + \frac{1}{2} \delta \(\rho\bb\) \cdot \bbprim - 2 \rho' \delta u\)\div \vv + 2\rho' \delta \(\frac{\pmat}{\rho}\) : \grad \vv \rangle  \\
&+ \langle \( - 2  \rho \vv \cdot \delta \bb - \rho \bb \cdot \delta \vv + \delta (\rho \vv) \cdot \bb   \) \divprim \bbprim +\( 2  \rho' \vvprim \cdot \delta \bb + \rho' \bbprim \cdot \delta \vv - \delta (\rho \vv) \cdot \bbprim   \)\div \bb  \rangle  \\
&+ \langle\(\delta \rho \frac{\ptot}{\rho} \cdot \vv -  \rho \delta \(\frac{\ptot}{\rho}\) \cdot \vv \)\cdot \frac{\gradprim \rho' }{\rho'} +\( \rho' \delta \(\frac{\ptot}{\rho} \)\cdot \vvprim - \delta \rho \frac{\ptotprim}{\rho'} \cdot \vvprim \) \cdot \frac{\grad \rho }{\rho} \rangle + \mathcal{F} + \mathcal{F'} + \mathcal{D}_k + \mathcal{D}'_k + \mathcal{D}_m + \mathcal{D}'_m \, . \end{split}\label{eq:KHM}
\ea 
\end{widetext}

From this relation and following the usual assumptions used in fully developed homogeneous turbulence, namely infinite kinetic and magnetic Reynolds numbers, stationary state, balance between forcing (at the largest scales) and dissipation (at the smallest ones) \citep{kolmogorov_dissipation_1941, frisch_turbulence_1995, galtier_exact_2011}, we obtain the following exact law valid in the inertial range:
\begin{widetext}
\ba
-4 & \varepsilon^{\text{MHD}} &= \divl \bm{\mathcal{F}^{\text{MHD}}} + \mathcal{S}^{\text{MHD}} + \mathcal{S'}{}^{\text{MHD}} \text{ with } \nonumber \\
&&\left\{
    \begin{split}
    \bm{\mathcal{F}^{\text{MHD}}} &= \langle \(\delta \(\rho \vv\) \cdot \delta \vv  + \delta \(\rho \bb\) \cdot \delta \bb + 2 \delta \rho \delta u\) \delta \vv -\( \delta \(\rho \bb\) \cdot \delta \vv + \delta \(\rho \vv\) \cdot \delta \bb \) \delta \bb - \delta \rho \delta \(\frac{\ptot}{\rho}\)\cdot \delta \vv  \rangle \, ,\\
\mathcal{S}^{\text{MHD}}& = \langle \(\rho \vv \cdot \delta \vv  +\frac{1}{2} \rho \bb \cdot \delta \bb - \frac{1}{2}\bb \cdot \delta \(\rho \bb\) + 2 \rho \delta u\)\divprim \vvprim - 2\rho \delta \(\frac{\pmat}{\rho}\) : \gradprim \vvprim \rangle  \\
&+ \langle \( - 2  \rho \vv \cdot \delta \bb - \rho \bb \cdot \delta \vv + \delta (\rho \vv) \cdot \bb   \) \divprim \bbprim \rangle + \langle\(\delta \rho \frac{\ptot}{\rho} \cdot \vv -  \rho \delta \(\frac{\ptot}{\rho}\) \cdot \vv \)\cdot \frac{\gradprim \rho' }{\rho'} \rangle \, , \\
\mathcal{S'}{}^{\text{MHD}}&= \text{ conjugate} \(\mathcal{S}{}^{\text{MHD}}\)\, , 
\end{split}
\right. \label{eq:extL}
\ea 
\end{widetext}

where $\varepsilon^{\text{MHD}}$ is the classical mean energy dissipation rate by unit mass assumed to be equal to the injection rate due to the forcing, i.e. $\mathcal{F} + \mathcal{F'} \simeq 4 \varepsilon^{\text{MHD}}$, and to the cascade rate in the inertial range due to non-linearities. The exact law (\ref{eq:extL}) is the first main result of this paper. It is valid for any MHD flow with a (symmetric) pressure tensor when the heat flux is neglected.

As in other compressible exact laws, we can recognize the terms introduced by \cite{andres_alternative_2017}: $\mathcal{F}^{\text{MHD}}$ is the flux terms (increment derivative $\divl \langle \rangle$), $\mathcal{S}^{\text{MHD}}$ and its conjugate $\mathcal{S'}{}^{\text{MHD}}$ are generally known as source terms (see below about the physical meaning of this terminology) where terms in $\langle \div \vv \rangle$ reflect the role of velocity dilatation, terms in $\langle \div \bb \rangle$ involve the (compressible) Alfv\'en speed dilatation and terms in $\langle \grad \rho \rangle$ contain density dilatation. Note that some hybrid and the $\beta$-dependent terms introduced in \citep{andres_alternative_2017} are hidden in the new structure function $\langle \delta \rho \delta\Big( \frac{\ptot}{\rho}\Big) \cdot \delta \vv \rangle$ and the terms in $\langle \grad \rho \rangle$.

\subsection{Extension to pressure anisotropic Hall-MHD}\label{hall}

The extension of the previous MHD model to Hall-MHD flows can be readily obtained by noticing that the only change to the original model is to introduce the Hall term in equation (\ref{eq:sys4}), while the internal energy equation remains unchanged. Therefore, the changes to the exact law (\ref{eq:extL}) will occur through the sole terms that depend on the current density, which were already derived in \cite{andres_exact_2018} for compressible isothermal MHD, without impacting pressure terms. The final exact law for Hall-MHD thus writes 
\begin{widetext}
\ba
-4 \varepsilon^{\text{HMHD}} &=& -4 \varepsilon^{\text{MHD}} + 2 d_i \divl \langle \overline{\rho \jjc \times \bb} \times \delta \bb - \delta(\jjc \times \bb)\times \overline{\rho \bb}\rangle \nonumber\\ 
&& -  \frac{d_i}{2} \langle \( \delta \rho \bbprim \cdot\bb \)\div \jjc  - \(\delta \rho \bbprim \cdot\bb \) \divprim \jjcprim\rangle + d_i \langle  \( \delta\rho \jjc \cdot \bbprim \)\div \bb - \( \delta \rho \jjcprim \cdot \bb\)\divprim \bbprim \rangle \, . \label{eq:extLHall}
\ea 
\end{widetext}
where $\varepsilon^{\text{MHD}}$ is given by equation (\ref{eq:extL}),  $d_i$ is the ion inertial length, and $\jj =\rho \jjc$ is the current density in Alfvénic units. 

\subsection{In the isotropic pressure case}

When considering a (total) scalar pressure $\pmat = P \id$ the MHD exact law (\ref{eq:extL}) takes the form:
\begin{widetext}
\ba
-4 & \varepsilon^{\text{MHD}} &= \divl \bm{\mathcal{F}^{\text{MHD}}} + \mathcal{S}^{\text{MHD}} + \mathcal{S'}{}^{\text{MHD}} \text{ with } \nonumber \\
&&\left\{
    \begin{split}
    \bm{\mathcal{F}^{\text{MHD}}} &= \langle \(\delta \(\rho \vv\) \cdot \delta \vv  + \delta \(\rho \bb\) \cdot \delta \bb + 2 \delta \rho \delta u\) \delta \vv -\( \delta \(\rho \bb\) \cdot \delta \vv + \delta \(\rho \vv\) \cdot \delta \bb \) \delta \bb - \delta \rho \delta \(\frac{P_*}{\rho}\)\delta \vv  \rangle \\
\mathcal{S}^{\text{MHD}}& = \langle \(\rho \vv \cdot \delta \vv  +\frac{1}{2} \rho \bb \cdot \delta \bb - \frac{1}{2}\bb \cdot \delta \(\rho \bb\) + 2 \rho \delta u - 2\rho \delta \(\frac{P}{\rho}\)\)\divprim \vvprim  \rangle  \\
&+ \langle \( - 2  \rho \vv \cdot \delta \bb - \rho \bb \cdot \delta \vv + \delta (\rho \vv) \cdot \bb   \) \divprim \bbprim \rangle + \langle\(\delta \rho \frac{P_*}{\rho} \cdot \vv -  \rho \delta \(\frac{P_*}{\rho}\) \cdot \vv \)\cdot \frac{\gradprim \rho' }{\rho'} \rangle \\
\mathcal{S'}{}^{\text{MHD}}&= \text{ conjugate} \(\mathcal{S}{}^{\text{MHD}}\)\, . 
\end{split}
\right. \label{eq:lim_sca}
\ea 
\end{widetext}

One can notice in relation (\ref{eq:lim_sca}) the presence of a new flux term that was not recognized as such in the previous models derived for scalar pressure \citep{banerjee_exact_2013,andres_alternative_2017,simon_general_2021}: $\divl \langle - \delta \rho \delta \(\frac{\ptot}{\rho}\) \cdot \delta \vv \rangle=\divl \langle - \delta \rho \delta \(\frac{P_*}{\rho}\) \delta \vv \rangle $. Using the first law of thermodynamics  $\rho^2 \grad u = P \grad \rho$, one can write the term in $\gradprim \rho'$ of $\mathcal{S}{}^{\text{MHD}}$ of equation (\ref{eq:lim_sca}) as (the same holds for its conjugate)
\begin{widetext}
\be
\langle \(\delta \(\rho\) \frac{P_*}{\rho} \vv -  \rho \delta \(\frac{P_*}{\rho}\) \vv \) \cdot \frac{\gradprim \rho' }{\rho'}\rangle =\langle \( \delta \(\frac{\rho^2}{P}\) \frac{P_*}{\rho} \vv -  \delta \( \frac{P_*}{P}\) \rho \vv \)\cdot  \gradprim u'\rangle =\langle \frac{P_*}{\rho} \vv \cdot \gradprim \rho' - \frac{P'_*}{P'} \divprim \(\rho u' \vv\)\rangle \, , \label{eq:betaterm}
\ee 
\end{widetext}
It is worth noting that the $\beta$-dependent term introduced by \cite{andres_alternative_2017} is hidden in this line since $P_*/P = 1 + P_M/P = 1+\beta^{-1}$. After some other manipulations, we recover the general exact law for isentropic flows derived in \cite{simon_general_2021}:
\begin{widetext}
\ba
- 4 \varepsilon^{\text{MHD}}
    &=& \divl \langle \delta (\rho \vv ) \cdot \delta \vv \delta \vv + \delta (\rho \bb ) \cdot \delta \bb \delta \vv + 2 \delta \rho \delta u \delta \vv - \delta (\rho \bb ) \cdot \delta \vv \delta \bb - \delta (\rho \vv ) \cdot \delta \bb \delta \bb \rangle \nonumber \\
    &&+ \divl  \langle  \(1+\frac{ \rho' }{\rho}\) (P + P_M) \vv' - \(1+\frac{ \rho }{\rho'}\)(P' + P'_M) \vv + \rho' u \vv' - \rho u' \vv \rangle \nonumber \\
    &&+ \langle (\divprim \vv')\(\rho \vv \cdot \delta \vv + \rho \bb \cdot \delta \bb - \frac{1}{2} \rho' \bb' \cdot \bb - \frac{1}{2} \rho  \bb \cdot \bb' + 2\rho \(\delta u - \frac{P'}{\rho'}\)\) \rangle \nonumber \\
    &&+ \langle (\div \vv)\(- \rho' \vv' \cdot \delta \vv - \rho' \bb' \cdot \delta \bb - \frac{1}{2} \rho  \bb \cdot \bb' - \frac{1}{2} \rho'  \bb' \cdot \bb - 2 \rho' \(\delta u + \frac{P}{\rho}\) \) \rangle \nonumber \\
    &&-  \langle (\divprim \bb')(2\rho \vv \cdot \delta \bb - \rho' \vv' \cdot \bb + \rho \bb \cdot \vv' ) - (\div \bb)(2 \rho' \vv' \cdot \delta \bb + \rho \vv \cdot \bb' - \rho' \bb' \cdot \vv ) \rangle \nonumber \\
    &&- \langle \frac{P'_M}{P'} \divprim (\rho u' \vv) + \frac{P_M}{P} \div (\rho' u \vv') \rangle \, . \label{eq:isent}
\ea 
\end{widetext}
It is worth recalling that this exact law is an extension of all scalar pressure models such as the isothermal and polytropic, which can be obtained by introducing the adequate state equation in relation (\ref{eq:isent}) (i.e., specifying the relation between the pressure $P$ and the density $\rho$) that are compatible with the isentropic hypothesis \citep{simon_general_2021}.

\section{Compressible MHD Exact law with a gyrotropic pressure}
The gyrotropic exact law can be readily obtained from relation (\ref{eq:extL}) by imposing the pressure tensor decomposition  $\pmat = \pperp \id + (\ppar - \pperp) \Bdir \Bdir$, with $\Bdir = \bb / |\bb|$ the magnetic field direction \citep{hunana_introductory_2019}. These definitions yield the following form of the total pressure $\ptot = \(\pperp + P_M\)  \id + (\ppar - \pperp) \Bdir \Bdir$. Using the tensor pressure equation, one can define the internal energy density as $ \rho u =  \frac{1}{2} \pmat:\id = \frac{1}{2} \ppar + \pperp $ \citep{hazeltine_local_2013,hunana_introductory_2019}. To highlight the terms in the exact law (\ref{eq:extL}) that can be linked to known (linear) plasma instabilities \citep{hunana_introductory_2019}, we further introduce the parameters $\beta_{\parallel} = \frac{\ppar}{P_M}$ and $a_p = \frac{\pperp}{\ppar} = T_\perp/T_\parallel$. Injecting these relations in equation (\ref{eq:extL}) yields the new gyrotropic-MHD exact law, which is the second main result of this paper:
\begin{widetext}
\ba
-4 & \varepsilon^{\text{GYR}} &= \divl \bm{\mathcal{F}^{\text{GYR}}} + \mathcal{S}^{\text{GYR}} + \mathcal{S'}{}^{\text{GYR}} \text{ with } \nonumber \\
&&\left\{
    \begin{split}
    \bm{\mathcal{F}^{\text{GYR}}} &= \langle \delta \(\rho \vv\) \cdot \delta \vv \delta \vv + \delta \(\rho \bb\) \cdot \delta \bb \delta \vv  - \delta \(\rho \bb\) \cdot \delta \vv \delta \bb - \delta \(\rho \vv\) \cdot \delta \bb \delta \bb \rangle  \\
&+ \langle \delta \rho \delta \(\frac{\bb^2 }{2}\(\beta_{\parallel}[1+a_p]-1\)\)\delta \vv - \delta \rho \delta \(\frac{ \beta_{\parallel}}{2}[1-a_p]\bb \bb\)\cdot \delta \vv\rangle \, , \\
\mathcal{S}^{\text{GYR}} &= \langle \(\rho \vv \cdot \delta \vv  +\frac{1}{2} \rho \bb \cdot \delta \bb - \frac{1}{2}\bb \cdot \delta \(\rho \bb\) +\rho \delta \(\frac{\bb^2 \beta_{\parallel}}{2} \)\)\divprim \vvprim - \rho \delta \( \beta_{\parallel}[1-a_p]\bb \bb\) : \gradprim \vvprim \rangle \\
&+\langle \( - 2  \rho \vv \cdot \delta \bb - \rho \bb \cdot \delta \vv + \delta (\rho \vv) \cdot \bb \) \divprim \bbprim\rangle \\
&+ \langle \(\(\delta \rho\) \frac{\bb^2}{2}[a_p \beta_{\parallel}+1]\vv -  \rho \delta \(\frac{\bb^2}{2}[a_p \beta_{\parallel}+1]\)  \vv  \)\cdot \frac{\gradprim \rho' }{\rho'} \rangle  \\
&+ \langle \(\(\delta \rho\) \frac{ \beta_{\parallel}}{2}[1-a_p]\bb \bb \cdot \vv -  \rho \delta \(\frac{ \beta_{\parallel}}{2}[1-a_p]\bb \bb\) \cdot \vv \)\cdot \frac{\gradprim \rho' }{\rho'}  \rangle \, ,  \\
\mathcal{S'}{}^{\text{GYR}}&= \text{ conjugate} \(\mathcal{S}{}^{\text{GYR}}\)\, . 
\end{split}
\right. \label{eq:extLgyr}
\ea 
\end{widetext}

Equation (\ref{eq:extLgyr}) shows the presence of new terms brought in by pressure anisotropy, which reveals how the turbulent cascade can be connected to the plasma instability conditions. For instance, the terms proportional to $ 1-a_p$ will have either positive or negative contribution to the cascade rate depending on the stability condition $a_p>1$ or $a_p<1$. In case of a positive (resp. negative) contribution to the cascade rate, pressure anisotropy can be seen as a source of ``free energy" (resp. a sink) that can reinforce (resp. diminish) the turbulence cascade. Furthermore, if the pressure anisotropy terms dominate the cascade then the instability would impact both the value of the energy cascade rate and its ``sense" (direct vs. inverse). Equation (\ref{eq:extLgyr}), which can be used on simulation and spacecraft data, may thus provide a solid theoretical explanation of the results reported in \cite{osman_anisotropic_2011,hadid_compressible_2018} and to the overall prominent role of the instabilities (non necessarily linear) in controlling part of the dynamics in astrophysical plasmas \citep{hellinger_solar_2006,bale_magnetic_2009,schekochihin_astrophysical_2009,kunz_firehose_2014}. 

In relation (\ref{eq:extLgyr}) the parameters $\beta_{\parallel}$ and $a_p$ that depend on the pressure components $\ppar$ and $\pperp$ are not yet determined since this relation derives from the internal energy equation (\ref{eq:sys3}), which constrains the sum of the two pressure components but not the individual ones. The latter can be determined by further introducing any closure equation compatible with the definition of the internal energy $\rho u = \frac{1}{2} \pmat:\id$ for each pressure component as done in the CGL-MHD theory.

We note finally that the Hall correction derived in section \ref{hall} remains valid  with this gyrotropic version of the exact law.

\subsection{Exact law for the CGL-MHD system}
The CGL-MHD closure equations written in their conservative form \citep{hunana_introductory_2019} read
\be \label{close_cgl}
\frac{d}{dt}\(\frac{\ppar B^2}{\rho^3}\)  = 0 \, \text{, and }\frac{d}{dt}\(\frac{\pperp}{\rho B}\)  = 0 \, ,
\ee 
where $d/dt$ is the total time derivative. Equations (\ref{close_cgl}) lead to the integrated form of the pressures and, consequently, to the forms of the parameters $\beta_{\parallel} = 2 C_{\parallel} \frac{ \rho}{\bb^4}$ and $a_p = C_{p} \frac{ |\bb|^3}{\rho^{1/2}}$, where the constants $C_{\parallel}$ and $C_{p}$ guaranty the homogeneity. Injecting these integrated relations in equation (\ref{eq:extLgyr}) yields the new CGL-MHD exact law, which is the third result of this paper:
\begin{widetext}
\ba
-4 & \varepsilon^{\text{CGL}}& = \divl \bm{\mathcal{F}^{\text{CGL}}} + \mathcal{S}^{\text{CGL}} + \mathcal{S'}{}^{\text{CGL}} \text{ with } \nonumber \\
&&\left\{
\begin{split}
\bm{\mathcal{F}^{\text{CGL}}} &= \langle \delta \(\rho \vv\) \cdot \delta \vv \delta \vv + \delta \(\rho \bb\) \cdot \delta \bb \delta \vv  - \delta \(\rho \bb\) \cdot \delta \vv \delta \bb - \delta \(\rho \vv\) \cdot \delta \bb \delta \bb \rangle  \\
&+ \langle \delta \rho \delta \(\frac{\bb^2 }{2}\( 2 C_{\parallel} \frac{ \rho}{\bb^4}[1+C_{p} \frac{ |\bb|^3}{\rho^{1/2}}]-1\)\)\delta \vv - \delta \rho \delta \( C_{\parallel} \frac{ \rho}{\bb^4}[1-C_{p} \frac{ |\bb|^3}{\rho^{1/2}}]\bb \bb\)\cdot \delta \vv\rangle \, , \\
\mathcal{S}^{\text{CGL}} &= \langle \(\rho \vv \cdot \delta \vv  +\frac{1}{2} \rho \bb \cdot \delta \bb - \frac{1}{2}\bb \cdot \delta \(\rho \bb\) +\rho \delta \(\frac{ C_{\parallel} \rho}{\bb^2} \)\)\divprim \vvprim  \rangle \\
&-\langle 2 \rho \delta \( C_{\parallel} \frac{ \rho}{\bb^4}[1-C_{p} \frac{ |\bb|^3}{\rho^{1/2}}]\bb \bb\) : \gradprim \vvprim + \( 2  \rho \vv \cdot \delta \bb + \rho \bb \cdot \delta \vv - \delta (\rho \vv) \cdot \bb \) \divprim \bbprim\rangle \\
&+ \langle \(\(\delta \rho\) \frac{\bb^2}{2}[2 C_{p} C_{\parallel} \frac{\rho^{1/2}}{|\bb|}+1] \vv -  \rho \delta \(\frac{\bb^2}{2}[2 C_{p} C_{\parallel} \frac{\rho^{1/2}}{|\bb|}+1]\)  \vv  \)\cdot \frac{\gradprim \rho' }{\rho'} \rangle  \\
&+ \langle \(\(\delta \rho\) C_{\parallel} \frac{ \rho}{\bb^4}[1-C_{p} \frac{ |\bb|^3}{\rho^{1/2}}]\bb \bb \cdot \vv -  \rho \delta \( C_{\parallel} \frac{ \rho}{\bb^4}[1-C_{p} \frac{ |\bb|^3}{\rho^{1/2}}]\bb \bb\) \cdot \vv \)\cdot \frac{\gradprim \rho' }{\rho'}  \rangle \, ,  \\
\mathcal{S'}{}^{\text{CGL}}&= \text{ conjugate} \(\mathcal{S}{}^{\text{CGL}}\)\, . \end{split}
\right. \label{eq:extLcgl}
\ea 
\end{widetext}

In the isotropic limit $\ppar = \pperp$ one finds the adiabatic (mono-atomic) case with a polytropic index $\gamma = 5/3$ and $\rho u = 3P/2$. Note that in the CGL-Hall-MHD the pressure equations do not write in a conservative form as those of the CGL-MHD (see equation (\ref{close_cgl})) \citep{hunana_introductory_2019}. This prevents us from obtaining a reduced form of the exact law for the CGL-Hall-MHD as that of the CGL-MHD. Nevertheless, the exact law (\ref{eq:extLHall}) is applicable to any CGL-Hall-MHD simulation data since the closure equations of the latter are compatible with the internal energy (equation (\ref{eq:sys3})) used to derive the law (\ref{eq:extLHall}) above.

\subsection{The incompressible MHD with a gyrotropic pressure: a generalization of the Politano and Pouquet's law}
In the incompressible limit, i.e. $\rho = \rho_0$ and $\div \vv = 0$, equation (\ref{eq:extLgyr}) becomes :
\be
\label{eq:extLinc} 4 \varepsilon^{IGYR} = 4 \varepsilon^{PP98} + \rho_0 \langle  \delta \( \beta_{\parallel}[1-a_p]\bb \bb\) : \delta ( \grad \vv)  \rangle \,,
\ee 
where $\varepsilon^{IGYR}$ stands for the cascade rate of incompressible gyrotropic model and $ -4 \varepsilon^{PP98} = \rho_0 \divl \langle \(\delta  \vv \cdot \delta \vv  + \delta \bb \cdot \delta \bb \) \delta \vv - 2 \delta \bb \cdot \delta \vv \delta \bb  \rangle$ is the so-called Politano and Pouquet's law \citep{politano_von_1998}, hereafter PP98, Interestingly, we evidence in equation (\ref{eq:extLinc}) the presence of a new {\it source} term brought in by the anisotropy of the pressure tensor, which is written as a contraction of two increment tensors. Equation (\ref{eq:extLinc}) is the fourth result of this paper. It generalizes PP98 to incompressible plasmas with a gyrotropic pressure and the notion of source terms. Indeed, so far the terminology of ``source" terms introduced in \cite{galtier_exact_2011} reflects compression (resp. dilatation) of the plasma that can sustain (resp. oppose) the cascade in the  inertial range \cite{ferrand_compressible_2020}. Here we evidence a new source term in the incompressible gyrotropic limit that is not tied to plasma contraction/dilatation, but to pressure anisotropy. It reflects the exchange between the no longer-conserved internal energy (unlike in incompressible pressure-isotropic flows \citep{simon_general_2021}) with the sum of the magnetic and kinetic energies as can be seen in equation (\ref{eq:sys3}) where we have $- \frac{\pmat}{\rho_0} : \grad \vv \neq 0$. This leads us to propose the following generalization of the notion of source: for compressible isentropic flows with a gyrotropic pressure tensor, the cascade of the kinetic and magnetic energies can be opposed/sustained by compression/dilatation of the fluid {\it and} by pressure anisotropy, the latter being relevant even in incompressible flows. For weakly compressible plasmas (e.g., SW), this result implies that the first order correction to the PP98 law would not come from density fluctuations, but rather from (incompressible) pressure anisotropy.

Similarly to the compressible gyrotropic case discussed above, the parameters $\beta_{\parallel}$ and $a_p$ remain undetermined. To determine the pressure components $\ppar$ and $\pperp$ the internal energy equation (\ref{eq:sys3}) (with now $\rho=\rho_0$) is complemented by a new equation coming from imposing the incompressiblity condition $\div \vv = 0$ on the momentum equation (\ref{eq:sys2}) (with $\rho=\rho_0$, and ${\bf d_k}= {\bf f}=0$ for simplicity), as done for incompressible isotropic hydrodynamics \citep{frisch_turbulence_1995} or Hall-MHD \citep{sahraoui_waves_2007}. This yields the generalized pressure balance equation for incompressible gyrotropic pressure tensor, namely,
\be
\label{eq:balance} \div\div (\rho \bb \bb-\rho \vv \vv-\ptot)=0 \, , \\
\ee 
Solving equations (\ref{eq:sys3}) (with $\rho=\rho_0$) and (\ref{eq:balance}) allows one to close the new incompressible gyrotropic MHD system proposed here and to self-consistently determine $\ppar$ and $\pperp$. However, for nearly incompressible plasmas such as the SW, the exact law (\ref{eq:extLinc}) can be directly applied to spacecraft data when $\ppar$ and $\pperp$ are accessible to measurements assuming equation (\ref{eq:balance}) to hold, as it has been done in all previous observational studies that used the PP98 model (assuming a scalar pressure). 

Note finally that the new model of incompressible gyrotropic (whose exact law is given by equation (\ref{eq:extLinc})) admits the oblique firehose instability as a linear solution, which is the unstable version of the known shear Alfv\'en mode \citep{hunana_introductory_2019}.

\section{Conclusion}
We derived new general exact laws for homogeneous MHD and Hall-MHD turbulent flows that go beyond the pressure isotropy assumption, which make them more realistic to study strong turbulence in magnetized plasmas. By considering the specific case of a CGL closure, we showed that the new law involves new flux and source terms that potentially can reflect the impact of plasma instabilities on the turbulent cascade. In the limit of incompressible MHD with a gyrotropic pressure we provided a generalization of the Politano and Pouquet's law \citep{politano_dynamical_1998} to pressure anisotropic plasmas, where a new incompressible source term is revealed and highlights a fundamental difference between pressure isotropic and anisotropic plasmas: internal energy is not conserved in the latter and pressure anisotropy can act as a source of free-energy to supply the turbulent cascade with an additional energy. This work thus paves the road to new and more rigorous (albeit fluid) studies of the interplay between turbulent (fluid) cascade and plasma instabilities, both in numerical simulations and spacecraft observations when the full pressure tensor is accessible to measurements.

{\bf Acknowledgement.---} PS is funded by a DIM-ACAV+ Doctoral fellowship. FS thanks T. Passot for fruitful discussions.


\begin{thebibliography}{61}%
\makeatletter
\providecommand \@ifxundefined [1]{%
 \@ifx{#1\undefined}
}%
\providecommand \@ifnum [1]{%
 \ifnum #1\expandafter \@firstoftwo
 \else \expandafter \@secondoftwo
 \fi
}%
\providecommand \@ifx [1]{%
 \ifx #1\expandafter \@firstoftwo
 \else \expandafter \@secondoftwo
 \fi
}%
\providecommand \natexlab [1]{#1}%
\providecommand \enquote  [1]{``#1''}%
\providecommand \bibnamefont  [1]{#1}%
\providecommand \bibfnamefont [1]{#1}%
\providecommand \citenamefont [1]{#1}%
\providecommand \href@noop [0]{\@secondoftwo}%
\providecommand \href [0]{\begingroup \@sanitize@url \@href}%
\providecommand \@href[1]{\@@startlink{#1}\@@href}%
\providecommand \@@href[1]{\endgroup#1\@@endlink}%
\providecommand \@sanitize@url [0]{\catcode `\\12\catcode `\$12\catcode
  `\&12\catcode `\#12\catcode `\^12\catcode `\_12\catcode `\%12\relax}%
\providecommand \@@startlink[1]{}%
\providecommand \@@endlink[0]{}%
\providecommand \url  [0]{\begingroup\@sanitize@url \@url }%
\providecommand \@url [1]{\endgroup\@href {#1}{\urlprefix }}%
\providecommand \urlprefix  [0]{URL }%
\providecommand \Eprint [0]{\href }%
\providecommand \doibase [0]{https://doi.org/}%
\providecommand \selectlanguage [0]{\@gobble}%
\providecommand \bibinfo  [0]{\@secondoftwo}%
\providecommand \bibfield  [0]{\@secondoftwo}%
\providecommand \translation [1]{[#1]}%
\providecommand \BibitemOpen [0]{}%
\providecommand \bibitemStop [0]{}%
\providecommand \bibitemNoStop [0]{.\EOS\space}%
\providecommand \EOS [0]{\spacefactor3000\relax}%
\providecommand \BibitemShut  [1]{\csname bibitem#1\endcsname}%
\let\auto@bib@innerbib\@empty
\bibitem [{\citenamefont {von Karman}\ and\ \citenamefont
  {Howarth}(1938)}]{von_karman_statistical_1938}%
  \BibitemOpen
  \bibfield  {author} {\bibinfo {author} {\bibfnamefont {T.}~\bibnamefont {von
  Karman}}\ and\ \bibinfo {author} {\bibfnamefont {L.}~\bibnamefont
  {Howarth}},\ }\bibfield  {title} {\bibinfo {title} {On the {Statistical}
  {Theory} of {Isotropic} {Turbulence}},\ }\href
  {https://doi.org/10.1098/rspa.1938.0013} {\bibfield  {journal} {\bibinfo
  {journal} {Proc. R. Soc. Lond. A}\ }\textbf {\bibinfo {volume} {164}},\
  \bibinfo {pages} {192} (\bibinfo {year} {1938})}\BibitemShut {NoStop}%
\bibitem [{\citenamefont {Monin}\ and\ \citenamefont
  {Jaglom}(1971)}]{monin_statistical_1971}%
  \BibitemOpen
  \bibfield  {author} {\bibinfo {author} {\bibfnamefont {A.~S.}\ \bibnamefont
  {Monin}}\ and\ \bibinfo {author} {\bibfnamefont {A.~M.}\ \bibnamefont
  {Jaglom}},\ }\href@noop {} {\emph {\bibinfo {title} {Statistical fluid
  mechanics: mechanics of turbulence}}},\ \bibinfo {edition} {english ed.
  updated, augmented and rev}\ ed.,\ Vol.~\bibinfo {volume} {1}\ (\bibinfo
  {publisher} {MIT Press},\ \bibinfo {address} {Cambridge, Mass},\ \bibinfo
  {year} {1971})\BibitemShut {NoStop}%
\bibitem [{\citenamefont {Monin}\ and\ \citenamefont
  {Jaglom}(1975)}]{monin_statistical_1975}%
  \BibitemOpen
  \bibfield  {author} {\bibinfo {author} {\bibfnamefont {A.~S.}\ \bibnamefont
  {Monin}}\ and\ \bibinfo {author} {\bibfnamefont {A.~M.}\ \bibnamefont
  {Jaglom}},\ }\href@noop {} {\emph {\bibinfo {title} {Statistical fluid
  mechanics: mechanics of turbulence}}},\ \bibinfo {edition} {2nd}\ ed.,\
  Vol.~\bibinfo {volume} {2}\ (\bibinfo  {publisher} {MIT Press},\ \bibinfo
  {address} {Cambridge},\ \bibinfo {year} {1975})\ \bibinfo {note} {oCLC:
  245996380}\BibitemShut {NoStop}%
\bibitem [{\citenamefont
  {Kolmogorov}(1991{\natexlab{a}})}]{kolmogorov_dissipation_1991}%
  \BibitemOpen
  \bibfield  {author} {\bibinfo {author} {\bibfnamefont {A.~N.}\ \bibnamefont
  {Kolmogorov}},\ }\bibfield  {title} {\bibinfo {title} {Dissipation of
  {Energy} in the {Locally} {Isotropic} {Turbulence}},\ }\href
  {https://doi.org/10.1098/rspa.1991.0076} {\bibfield  {journal} {\bibinfo
  {journal} {Proc. R. Soc. A}\ }\textbf {\bibinfo {volume} {434}},\ \bibinfo
  {pages} {15} (\bibinfo {year} {1991}{\natexlab{a}})}\BibitemShut {NoStop}%
\bibitem [{\citenamefont
  {Kolmogorov}(1991{\natexlab{b}})}]{kolmogorov_local_1991}%
  \BibitemOpen
  \bibfield  {author} {\bibinfo {author} {\bibnamefont {Kolmogorov}},\
  }\bibfield  {title} {\bibinfo {title} {The local structure of turbulence in
  incompressible viscous fluid for very large {Reynolds} numbers},\ }\href
  {https://doi.org/10.1098/rspa.1991.0075} {\bibfield  {journal} {\bibinfo
  {journal} {Proc. R. Soc. Lond. A}\ }\textbf {\bibinfo {volume} {434}},\
  \bibinfo {pages} {9} (\bibinfo {year} {1991}{\natexlab{b}})}\BibitemShut
  {NoStop}%
\bibitem [{\citenamefont {Frisch}(1995)}]{frisch_turbulence_1995}%
  \BibitemOpen
  \bibfield  {author} {\bibinfo {author} {\bibfnamefont {U.}~\bibnamefont
  {Frisch}},\ }\href@noop {} {\emph {\bibinfo {title} {Turbulence: {The}
  {Legacy} of {A}.{N}. {Kolmogorov}}}}\ (\bibinfo  {publisher} {CAMBRIDGE
  UNIVERSITY PRESS},\ \bibinfo {year} {1995})\BibitemShut {NoStop}%
\bibitem [{\citenamefont {Antonia}\ \emph {et~al.}(1997)\citenamefont
  {Antonia}, \citenamefont {Ould-Rouis}, \citenamefont {Anselmet},\ and\
  \citenamefont {Zhu}}]{antonia_analogy_1997}%
  \BibitemOpen
  \bibfield  {author} {\bibinfo {author} {\bibfnamefont {R.~A.}\ \bibnamefont
  {Antonia}}, \bibinfo {author} {\bibfnamefont {M.}~\bibnamefont {Ould-Rouis}},
  \bibinfo {author} {\bibfnamefont {F.}~\bibnamefont {Anselmet}},\ and\
  \bibinfo {author} {\bibfnamefont {Y.}~\bibnamefont {Zhu}},\ }\bibfield
  {title} {\bibinfo {title} {Analogy between predictions of {Kolmogorov} and
  {Yaglom}},\ }\href {https://doi.org/10.1017/S0022112096004090} {\bibfield
  {journal} {\bibinfo  {journal} {J. Fluid Mech.}\ }\textbf {\bibinfo {volume}
  {332}},\ \bibinfo {pages} {395} (\bibinfo {year} {1997})},\ \bibinfo {note}
  {publisher: Cambridge University Press}\BibitemShut {NoStop}%
\bibitem [{\citenamefont {Zakharov}\ \emph {et~al.}(1992)\citenamefont
  {Zakharov}, \citenamefont {L'vov},\ and\ \citenamefont
  {Falkovich}}]{zakharov_kolmogorov_1992}%
  \BibitemOpen
  \bibfield  {author} {\bibinfo {author} {\bibfnamefont {V.~E.}\ \bibnamefont
  {Zakharov}}, \bibinfo {author} {\bibfnamefont {V.~S.}\ \bibnamefont
  {L'vov}},\ and\ \bibinfo {author} {\bibfnamefont {G.}~\bibnamefont
  {Falkovich}},\ }\href {https://doi.org/10.1007/978-3-642-50052-7} {\emph
  {\bibinfo {title} {Kolmogorov {Spectra} of {Turbulence} {I}: {Wave}
  {Turbulence}}}},\ Springer {Series} in {Nonlinear} {Dynamics}\ (\bibinfo
  {publisher} {Springer-Verlag},\ \bibinfo {address} {Berlin Heidelberg},\
  \bibinfo {year} {1992})\BibitemShut {NoStop}%
\bibitem [{\citenamefont {Galtier}(2016)}]{galtier_introduction_2016}%
  \BibitemOpen
  \bibfield  {author} {\bibinfo {author} {\bibfnamefont {S.}~\bibnamefont
  {Galtier}},\ }\href {https://doi.org/10.1017/CBO9781316665961} {\emph
  {\bibinfo {title} {Introduction to {Modern} {Magnetohydrodynamics}}}}\
  (\bibinfo  {publisher} {Cambridge University Press},\ \bibinfo {address}
  {Cambridge},\ \bibinfo {year} {2016})\BibitemShut {NoStop}%
\bibitem [{\citenamefont {Galtier}(2021)}]{galtier_physique_2021}%
  \BibitemOpen
  \bibfield  {author} {\bibinfo {author} {\bibfnamefont {S.}~\bibnamefont
  {Galtier}},\ }\href@noop {} {\emph {\bibinfo {title} {Physique de la
  {Turbulence}: des tourbillons aux ondes}}}\ (\bibinfo  {publisher} {CNRS
  Editions / EDP Sciences (Savoirs Actuels)},\ \bibinfo {year}
  {2021})\BibitemShut {NoStop}%
\bibitem [{\citenamefont {Smith}\ \emph {et~al.}(2006)\citenamefont {Smith},
  \citenamefont {Hamilton}, \citenamefont {Vasquez},\ and\ \citenamefont
  {Leamon}}]{smith_dependence_2006}%
  \BibitemOpen
  \bibfield  {author} {\bibinfo {author} {\bibfnamefont {C.~W.}\ \bibnamefont
  {Smith}}, \bibinfo {author} {\bibfnamefont {K.}~\bibnamefont {Hamilton}},
  \bibinfo {author} {\bibfnamefont {B.~J.}\ \bibnamefont {Vasquez}},\ and\
  \bibinfo {author} {\bibfnamefont {R.~J.}\ \bibnamefont {Leamon}},\ }\bibfield
   {title} {\bibinfo {title} {Dependence of the {Dissipation} {Range}
  {Spectrum} of {Interplanetary} {Magnetic} {Fluctuationson} the {Rate} of
  {Energy} {Cascade}},\ }\href {https://doi.org/10.1086/506151} {\bibfield
  {journal} {\bibinfo  {journal} {ApJ}\ }\textbf {\bibinfo {volume} {645}},\
  \bibinfo {pages} {L85} (\bibinfo {year} {2006})},\ \bibinfo {note}
  {publisher: IOP Publishing}\BibitemShut {NoStop}%
\bibitem [{\citenamefont {Sorriso-Valvo}\ \emph {et~al.}(2007)\citenamefont
  {Sorriso-Valvo}, \citenamefont {Marino}, \citenamefont {Carbone},
  \citenamefont {Noullez}, \citenamefont {Lepreti}, \citenamefont {Veltri},
  \citenamefont {Bruno}, \citenamefont {Bavassano},\ and\ \citenamefont
  {Pietropaolo}}]{sorriso-valvo_observation_2007}%
  \BibitemOpen
  \bibfield  {author} {\bibinfo {author} {\bibfnamefont {L.}~\bibnamefont
  {Sorriso-Valvo}}, \bibinfo {author} {\bibfnamefont {R.}~\bibnamefont
  {Marino}}, \bibinfo {author} {\bibfnamefont {V.}~\bibnamefont {Carbone}},
  \bibinfo {author} {\bibfnamefont {A.}~\bibnamefont {Noullez}}, \bibinfo
  {author} {\bibfnamefont {F.}~\bibnamefont {Lepreti}}, \bibinfo {author}
  {\bibfnamefont {P.}~\bibnamefont {Veltri}}, \bibinfo {author} {\bibfnamefont
  {R.}~\bibnamefont {Bruno}}, \bibinfo {author} {\bibfnamefont
  {B.}~\bibnamefont {Bavassano}},\ and\ \bibinfo {author} {\bibfnamefont
  {E.}~\bibnamefont {Pietropaolo}},\ }\bibfield  {title} {\bibinfo {title}
  {Observation of {Inertial} {Energy} {Cascade} in {Interplanetary} {Space}
  {Plasma}},\ }\href {https://doi.org/10.1103/PhysRevLett.99.115001} {\bibfield
   {journal} {\bibinfo  {journal} {Phys. Rev. Lett.}\ }\textbf {\bibinfo
  {volume} {99}},\ \bibinfo {pages} {115001} (\bibinfo {year} {2007})},\
  \bibinfo {note} {publisher: American Physical Society}\BibitemShut {NoStop}%
\bibitem [{\citenamefont {Andrés}\ \emph {et~al.}(2019)\citenamefont
  {Andrés}, \citenamefont {Sahraoui}, \citenamefont {Galtier}, \citenamefont
  {Hadid}, \citenamefont {Ferrand},\ and\ \citenamefont
  {Huang}}]{andres_energy_2019}%
  \BibitemOpen
  \bibfield  {author} {\bibinfo {author} {\bibfnamefont {N.}~\bibnamefont
  {Andrés}}, \bibinfo {author} {\bibfnamefont {F.}~\bibnamefont {Sahraoui}},
  \bibinfo {author} {\bibfnamefont {S.}~\bibnamefont {Galtier}}, \bibinfo
  {author} {\bibfnamefont {L.~Z.}\ \bibnamefont {Hadid}}, \bibinfo {author}
  {\bibfnamefont {R.}~\bibnamefont {Ferrand}},\ and\ \bibinfo {author}
  {\bibfnamefont {S.~Y.}\ \bibnamefont {Huang}},\ }\bibfield  {title} {\bibinfo
  {title} {Energy cascade rate measured in a collisionless space plasma with
  {MMS} data and compressible {Hall} magnetohydrodynamic turbulence theory},\
  }\href {https://doi.org/10.1103/PhysRevLett.123.245101} {\bibfield  {journal}
  {\bibinfo  {journal} {Phys. Rev. Lett.}\ }\textbf {\bibinfo {volume} {123}},\
  \bibinfo {pages} {245101} (\bibinfo {year} {2019})},\ \bibinfo {note} {arXiv:
  1911.09749}\BibitemShut {NoStop}%
\bibitem [{\citenamefont {Sorriso-Valvo}\ \emph {et~al.}(2019)\citenamefont
  {Sorriso-Valvo}, \citenamefont {Catapano}, \citenamefont {Retinò},
  \citenamefont {Le~Contel}, \citenamefont {Perrone}, \citenamefont {Roberts},
  \citenamefont {Coburn}, \citenamefont {Panebianco}, \citenamefont
  {Valentini}, \citenamefont {Perri}, \citenamefont {Greco}, \citenamefont
  {Malara}, \citenamefont {Carbone}, \citenamefont {Veltri}, \citenamefont
  {Pezzi}, \citenamefont {Fraternale}, \citenamefont {Di~Mare}, \citenamefont
  {Marino}, \citenamefont {Giles}, \citenamefont {Moore}, \citenamefont
  {Russell}, \citenamefont {Torbert}, \citenamefont {Burch},\ and\
  \citenamefont {Khotyaintsev}}]{sorriso-valvo_turbulence-driven_2019}%
  \BibitemOpen
  \bibfield  {author} {\bibinfo {author} {\bibfnamefont {L.}~\bibnamefont
  {Sorriso-Valvo}}, \bibinfo {author} {\bibfnamefont {F.}~\bibnamefont
  {Catapano}}, \bibinfo {author} {\bibfnamefont {A.}~\bibnamefont {Retinò}},
  \bibinfo {author} {\bibfnamefont {O.}~\bibnamefont {Le~Contel}}, \bibinfo
  {author} {\bibfnamefont {D.}~\bibnamefont {Perrone}}, \bibinfo {author}
  {\bibfnamefont {O.~W.}\ \bibnamefont {Roberts}}, \bibinfo {author}
  {\bibfnamefont {J.~T.}\ \bibnamefont {Coburn}}, \bibinfo {author}
  {\bibfnamefont {V.}~\bibnamefont {Panebianco}}, \bibinfo {author}
  {\bibfnamefont {F.}~\bibnamefont {Valentini}}, \bibinfo {author}
  {\bibfnamefont {S.}~\bibnamefont {Perri}}, \bibinfo {author} {\bibfnamefont
  {A.}~\bibnamefont {Greco}}, \bibinfo {author} {\bibfnamefont
  {F.}~\bibnamefont {Malara}}, \bibinfo {author} {\bibfnamefont
  {V.}~\bibnamefont {Carbone}}, \bibinfo {author} {\bibfnamefont
  {P.}~\bibnamefont {Veltri}}, \bibinfo {author} {\bibfnamefont
  {O.}~\bibnamefont {Pezzi}}, \bibinfo {author} {\bibfnamefont
  {F.}~\bibnamefont {Fraternale}}, \bibinfo {author} {\bibfnamefont
  {F.}~\bibnamefont {Di~Mare}}, \bibinfo {author} {\bibfnamefont
  {R.}~\bibnamefont {Marino}}, \bibinfo {author} {\bibfnamefont
  {B.}~\bibnamefont {Giles}}, \bibinfo {author} {\bibfnamefont {T.~E.}\
  \bibnamefont {Moore}}, \bibinfo {author} {\bibfnamefont {C.~T.}\ \bibnamefont
  {Russell}}, \bibinfo {author} {\bibfnamefont {R.~B.}\ \bibnamefont
  {Torbert}}, \bibinfo {author} {\bibfnamefont {J.~L.}\ \bibnamefont {Burch}},\
  and\ \bibinfo {author} {\bibfnamefont {Y.~V.}\ \bibnamefont {Khotyaintsev}},\
  }\bibfield  {title} {\bibinfo {title} {Turbulence-{Driven} {Ion} {Beams} in
  the {Magnetospheric} {Kelvin}-{Helmholtz} {Instability}},\ }\href
  {https://doi.org/10.1103/PhysRevLett.122.035102} {\bibfield  {journal}
  {\bibinfo  {journal} {Phys. Rev. Lett.}\ }\textbf {\bibinfo {volume} {122}},\
  \bibinfo {pages} {035102} (\bibinfo {year} {2019})},\ \bibinfo {note}
  {publisher: American Physical Society}\BibitemShut {NoStop}%
\bibitem [{\citenamefont {Bandyopadhyay}\ \emph {et~al.}(2020)\citenamefont
  {Bandyopadhyay}, \citenamefont {Sorriso-Valvo}, \citenamefont {Chasapis},
  \citenamefont {Hellinger}, \citenamefont {Matthaeus}, \citenamefont
  {Verdini}, \citenamefont {Landi}, \citenamefont {Franci}, \citenamefont
  {Matteini}, \citenamefont {Giles}, \citenamefont {Gershman}, \citenamefont
  {Moore}, \citenamefont {Pollock}, \citenamefont {Russell}, \citenamefont
  {Strangeway}, \citenamefont {Torbert},\ and\ \citenamefont
  {Burch}}]{bandyopadhyay_situ_2020}%
  \BibitemOpen
  \bibfield  {author} {\bibinfo {author} {\bibfnamefont {R.}~\bibnamefont
  {Bandyopadhyay}}, \bibinfo {author} {\bibfnamefont {L.}~\bibnamefont
  {Sorriso-Valvo}}, \bibinfo {author} {\bibfnamefont {A.}~\bibnamefont
  {Chasapis}}, \bibinfo {author} {\bibfnamefont {P.}~\bibnamefont {Hellinger}},
  \bibinfo {author} {\bibfnamefont {W.~H.}\ \bibnamefont {Matthaeus}}, \bibinfo
  {author} {\bibfnamefont {A.}~\bibnamefont {Verdini}}, \bibinfo {author}
  {\bibfnamefont {S.}~\bibnamefont {Landi}}, \bibinfo {author} {\bibfnamefont
  {L.}~\bibnamefont {Franci}}, \bibinfo {author} {\bibfnamefont
  {L.}~\bibnamefont {Matteini}}, \bibinfo {author} {\bibfnamefont {B.~L.}\
  \bibnamefont {Giles}}, \bibinfo {author} {\bibfnamefont {D.~J.}\ \bibnamefont
  {Gershman}}, \bibinfo {author} {\bibfnamefont {T.~E.}\ \bibnamefont {Moore}},
  \bibinfo {author} {\bibfnamefont {C.~J.}\ \bibnamefont {Pollock}}, \bibinfo
  {author} {\bibfnamefont {C.~T.}\ \bibnamefont {Russell}}, \bibinfo {author}
  {\bibfnamefont {R.~J.}\ \bibnamefont {Strangeway}}, \bibinfo {author}
  {\bibfnamefont {R.~B.}\ \bibnamefont {Torbert}},\ and\ \bibinfo {author}
  {\bibfnamefont {J.~L.}\ \bibnamefont {Burch}},\ }\bibfield  {title} {\bibinfo
  {title} {In {Situ} {Observation} of {Hall} {Magnetohydrodynamic} {Cascade} in
  {Space} {Plasma}},\ }\href {https://doi.org/10.1103/PhysRevLett.124.225101}
  {\bibfield  {journal} {\bibinfo  {journal} {Phys. Rev. Lett.}\ }\textbf
  {\bibinfo {volume} {124}},\ \bibinfo {pages} {225101} (\bibinfo {year}
  {2020})},\ \bibinfo {note} {publisher: American Physical Society}\BibitemShut
  {NoStop}%
\bibitem [{\citenamefont {Quijia}\ \emph {et~al.}(2021)\citenamefont {Quijia},
  \citenamefont {Fraternale}, \citenamefont {Stawarz}, \citenamefont
  {Vásconez}, \citenamefont {Perri}, \citenamefont {Marino}, \citenamefont
  {Yordanova},\ and\ \citenamefont {Sorriso-Valvo}}]{quijia_comparing_2021}%
  \BibitemOpen
  \bibfield  {author} {\bibinfo {author} {\bibfnamefont {P.}~\bibnamefont
  {Quijia}}, \bibinfo {author} {\bibfnamefont {F.}~\bibnamefont {Fraternale}},
  \bibinfo {author} {\bibfnamefont {J.~E.}\ \bibnamefont {Stawarz}}, \bibinfo
  {author} {\bibfnamefont {C.~L.}\ \bibnamefont {Vásconez}}, \bibinfo {author}
  {\bibfnamefont {S.}~\bibnamefont {Perri}}, \bibinfo {author} {\bibfnamefont
  {R.}~\bibnamefont {Marino}}, \bibinfo {author} {\bibfnamefont
  {E.}~\bibnamefont {Yordanova}},\ and\ \bibinfo {author} {\bibfnamefont
  {L.}~\bibnamefont {Sorriso-Valvo}},\ }\bibfield  {title} {\bibinfo {title}
  {Comparing turbulence in a {Kelvin}–{Helmholtz} instability region across
  the terrestrial magnetopause},\ }\href
  {https://doi.org/10.1093/mnras/stab319} {\bibfield  {journal} {\bibinfo
  {journal} {Monthly Notices of the Royal Astronomical Society}\ }\textbf
  {\bibinfo {volume} {503}},\ \bibinfo {pages} {4815} (\bibinfo {year}
  {2021})},\ \bibinfo {note} {\_eprint:
  https://academic.oup.com/mnras/article-pdf/503/4/4815/37011806/stab319.pdf}\BibitemShut
  {NoStop}%
\bibitem [{\citenamefont {Politano}\ and\ \citenamefont
  {Pouquet}(1998{\natexlab{a}})}]{politano_dynamical_1998}%
  \BibitemOpen
  \bibfield  {author} {\bibinfo {author} {\bibfnamefont {H.}~\bibnamefont
  {Politano}}\ and\ \bibinfo {author} {\bibfnamefont {A.}~\bibnamefont
  {Pouquet}},\ }\bibfield  {title} {\bibinfo {title} {Dynamical length scales
  for turbulent magnetized flows},\ }\href {https://doi.org/10.1029/97GL03642}
  {\bibfield  {journal} {\bibinfo  {journal} {Geophys. Res. Lett.}\ }\textbf
  {\bibinfo {volume} {25}},\ \bibinfo {pages} {273} (\bibinfo {year}
  {1998}{\natexlab{a}})}\BibitemShut {NoStop}%
\bibitem [{\citenamefont {Politano}\ and\ \citenamefont
  {Pouquet}(1998{\natexlab{b}})}]{politano_von_1998}%
  \BibitemOpen
  \bibfield  {author} {\bibinfo {author} {\bibfnamefont {H.}~\bibnamefont
  {Politano}}\ and\ \bibinfo {author} {\bibfnamefont {A.}~\bibnamefont
  {Pouquet}},\ }\bibfield  {title} {\bibinfo {title} {von
  {Kármán}–{Howarth} equation for magnetohydrodynamics and its consequences
  on third-order longitudinal structure and correlation functions},\ }\href
  {https://doi.org/10.1103/PhysRevE.57.R21} {\bibfield  {journal} {\bibinfo
  {journal} {Phys. Rev. E}\ }\textbf {\bibinfo {volume} {57}},\ \bibinfo
  {pages} {R21} (\bibinfo {year} {1998}{\natexlab{b}})}\BibitemShut {NoStop}%
\bibitem [{\citenamefont {Galtier}(2008)}]{galtier_von_2008}%
  \BibitemOpen
  \bibfield  {author} {\bibinfo {author} {\bibfnamefont {S.}~\bibnamefont
  {Galtier}},\ }\bibfield  {title} {\bibinfo {title} {von
  {Kármán}–{Howarth} equations for {Hall} magnetohydrodynamic flows},\
  }\href {https://doi.org/10.1103/PhysRevE.77.015302} {\bibfield  {journal}
  {\bibinfo  {journal} {Phys. Rev. E}\ }\textbf {\bibinfo {volume} {77}},\
  \bibinfo {pages} {015302} (\bibinfo {year} {2008})}\BibitemShut {NoStop}%
\bibitem [{\citenamefont {Andrés}\ \emph
  {et~al.}(2016{\natexlab{a}})\citenamefont {Andrés}, \citenamefont
  {Galtier},\ and\ \citenamefont {Sahraoui}}]{andres_exact_2016}%
  \BibitemOpen
  \bibfield  {author} {\bibinfo {author} {\bibfnamefont {N.}~\bibnamefont
  {Andrés}}, \bibinfo {author} {\bibfnamefont {S.}~\bibnamefont {Galtier}},\
  and\ \bibinfo {author} {\bibfnamefont {F.}~\bibnamefont {Sahraoui}},\
  }\bibfield  {title} {\bibinfo {title} {Exact scaling laws for helical
  three-dimensional two-fluid turbulent plasmas},\ }\href
  {https://doi.org/10.1103/PhysRevE.94.063206} {\bibfield  {journal} {\bibinfo
  {journal} {Phys. Rev. E}\ }\textbf {\bibinfo {volume} {94}},\ \bibinfo
  {pages} {063206} (\bibinfo {year} {2016}{\natexlab{a}})}\BibitemShut
  {NoStop}%
\bibitem [{\citenamefont {Andrés}\ \emph
  {et~al.}(2016{\natexlab{b}})\citenamefont {Andrés}, \citenamefont {Mininni},
  \citenamefont {Dmitruk},\ and\ \citenamefont {Gómez}}]{andres_von_2016}%
  \BibitemOpen
  \bibfield  {author} {\bibinfo {author} {\bibfnamefont {N.}~\bibnamefont
  {Andrés}}, \bibinfo {author} {\bibfnamefont {P.~D.}\ \bibnamefont
  {Mininni}}, \bibinfo {author} {\bibfnamefont {P.}~\bibnamefont {Dmitruk}},\
  and\ \bibinfo {author} {\bibfnamefont {D.~O.}\ \bibnamefont {Gómez}},\
  }\bibfield  {title} {\bibinfo {title} {von {Kármán}–{Howarth} equation
  for three-dimensional two-fluid plasmas},\ }\href
  {https://doi.org/10.1103/PhysRevE.93.063202} {\bibfield  {journal} {\bibinfo
  {journal} {Phys. Rev. E}\ }\textbf {\bibinfo {volume} {93}},\ \bibinfo
  {pages} {063202} (\bibinfo {year} {2016}{\natexlab{b}})}\BibitemShut
  {NoStop}%
\bibitem [{\citenamefont {Banerjee}\ and\ \citenamefont
  {Galtier}(2017)}]{banerjee_alternative_2017}%
  \BibitemOpen
  \bibfield  {author} {\bibinfo {author} {\bibfnamefont {S.}~\bibnamefont
  {Banerjee}}\ and\ \bibinfo {author} {\bibfnamefont {S.}~\bibnamefont
  {Galtier}},\ }\bibfield  {title} {\bibinfo {title} {An alternative
  formulation for exact scaling relations in hydrodynamic and
  magnetohydrodynamic turbulence},\ }\href
  {https://doi.org/10.1088/1751-8113/50/1/015501} {\bibfield  {journal}
  {\bibinfo  {journal} {J. Phys. A: Math. Theor.}\ }\textbf {\bibinfo {volume}
  {50}},\ \bibinfo {pages} {015501} (\bibinfo {year} {2017})}\BibitemShut
  {NoStop}%
\bibitem [{\citenamefont {Hellinger}\ \emph {et~al.}(2018)\citenamefont
  {Hellinger}, \citenamefont {Verdini}, \citenamefont {Landi}, \citenamefont
  {Franci},\ and\ \citenamefont {Matteini}}]{hellinger_von_2018}%
  \BibitemOpen
  \bibfield  {author} {\bibinfo {author} {\bibfnamefont {P.}~\bibnamefont
  {Hellinger}}, \bibinfo {author} {\bibfnamefont {A.}~\bibnamefont {Verdini}},
  \bibinfo {author} {\bibfnamefont {S.}~\bibnamefont {Landi}}, \bibinfo
  {author} {\bibfnamefont {L.}~\bibnamefont {Franci}},\ and\ \bibinfo {author}
  {\bibfnamefont {L.}~\bibnamefont {Matteini}},\ }\bibfield  {title} {\bibinfo
  {title} {von {Kármán}–{Howarth} {Equation} for {Hall}
  {Magnetohydrodynamics}: {Hybrid} {Simulations}},\ }\href
  {https://doi.org/10.3847/2041-8213/aabc06} {\bibfield  {journal} {\bibinfo
  {journal} {The Astrophysical Journal}\ }\textbf {\bibinfo {volume} {857}},\
  \bibinfo {pages} {L19} (\bibinfo {year} {2018})},\ \bibinfo {note}
  {publisher: American Astronomical Society}\BibitemShut {NoStop}%
\bibitem [{\citenamefont {Ferrand}\ \emph {et~al.}(2019)\citenamefont
  {Ferrand}, \citenamefont {Galtier}, \citenamefont {Sahraoui}, \citenamefont
  {Meyrand}, \citenamefont {Andrés},\ and\ \citenamefont
  {Banerjee}}]{ferrand_exact_2019}%
  \BibitemOpen
  \bibfield  {author} {\bibinfo {author} {\bibfnamefont {R.}~\bibnamefont
  {Ferrand}}, \bibinfo {author} {\bibfnamefont {S.}~\bibnamefont {Galtier}},
  \bibinfo {author} {\bibfnamefont {F.}~\bibnamefont {Sahraoui}}, \bibinfo
  {author} {\bibfnamefont {R.}~\bibnamefont {Meyrand}}, \bibinfo {author}
  {\bibfnamefont {N.}~\bibnamefont {Andrés}},\ and\ \bibinfo {author}
  {\bibfnamefont {S.}~\bibnamefont {Banerjee}},\ }\bibfield  {title} {\bibinfo
  {title} {On {Exact} {Laws} in {Incompressible} {Hall} {Magnetohydrodynamic}
  {Turbulence}},\ }\href {https://doi.org/10.3847/1538-4357/ab2be9} {\bibfield
  {journal} {\bibinfo  {journal} {ApJ}\ }\textbf {\bibinfo {volume} {881}},\
  \bibinfo {pages} {50} (\bibinfo {year} {2019})}\BibitemShut {NoStop}%
\bibitem [{\citenamefont {Galtier}\ and\ \citenamefont
  {Banerjee}(2011)}]{galtier_exact_2011}%
  \BibitemOpen
  \bibfield  {author} {\bibinfo {author} {\bibfnamefont {S.}~\bibnamefont
  {Galtier}}\ and\ \bibinfo {author} {\bibfnamefont {S.}~\bibnamefont
  {Banerjee}},\ }\bibfield  {title} {\bibinfo {title} {Exact {Relation} for
  {Correlation} {Functions} in {Compressible} {Isothermal} {Turbulence}},\
  }\href {https://doi.org/10.1103/PhysRevLett.107.134501} {\bibfield  {journal}
  {\bibinfo  {journal} {Phys. Rev. Lett.}\ }\textbf {\bibinfo {volume} {107}},\
  \bibinfo {pages} {134501} (\bibinfo {year} {2011})}\BibitemShut {NoStop}%
\bibitem [{\citenamefont {Banerjee}\ and\ \citenamefont
  {Galtier}(2013)}]{banerjee_exact_2013}%
  \BibitemOpen
  \bibfield  {author} {\bibinfo {author} {\bibfnamefont {S.}~\bibnamefont
  {Banerjee}}\ and\ \bibinfo {author} {\bibfnamefont {S.}~\bibnamefont
  {Galtier}},\ }\bibfield  {title} {\bibinfo {title} {Exact relation with
  two-point correlation functions and phenomenological approach for
  compressible magnetohydrodynamic turbulence},\ }\href@noop {} {\bibfield
  {journal} {\bibinfo  {journal} {Phys. Rev. E}\ }\textbf {\bibinfo {volume}
  {87}},\ \bibinfo {pages} {7} (\bibinfo {year} {2013})}\BibitemShut {NoStop}%
\bibitem [{\citenamefont {Banerjee}\ and\ \citenamefont
  {Galtier}(2014)}]{banerjee_kolmogorov-like_2014}%
  \BibitemOpen
  \bibfield  {author} {\bibinfo {author} {\bibfnamefont {S.}~\bibnamefont
  {Banerjee}}\ and\ \bibinfo {author} {\bibfnamefont {S.}~\bibnamefont
  {Galtier}},\ }\bibfield  {title} {\bibinfo {title} {A {Kolmogorov}-like exact
  relation for compressible polytropic turbulence},\ }\href
  {https://doi.org/10.1017/jfm.2013.657} {\bibfield  {journal} {\bibinfo
  {journal} {J. Fluid Mech.}\ }\textbf {\bibinfo {volume} {742}},\ \bibinfo
  {pages} {230} (\bibinfo {year} {2014})}\BibitemShut {NoStop}%
\bibitem [{\citenamefont {Banerjee}\ \emph {et~al.}(2016)\citenamefont
  {Banerjee}, \citenamefont {Hadid}, \citenamefont {Sahraoui},\ and\
  \citenamefont {Galtier}}]{banerjee_scaling_2016}%
  \BibitemOpen
  \bibfield  {author} {\bibinfo {author} {\bibfnamefont {S.}~\bibnamefont
  {Banerjee}}, \bibinfo {author} {\bibfnamefont {L.~Z.}\ \bibnamefont {Hadid}},
  \bibinfo {author} {\bibfnamefont {F.}~\bibnamefont {Sahraoui}},\ and\
  \bibinfo {author} {\bibfnamefont {S.}~\bibnamefont {Galtier}},\ }\bibfield
  {title} {\bibinfo {title} {Scaling of compressible magnetohydrodynamic
  turbulence in the fast solar wind},\ }\href
  {https://doi.org/10.3847/2041-8205/829/2/L27} {\bibfield  {journal} {\bibinfo
   {journal} {ApJL}\ }\textbf {\bibinfo {volume} {829}},\ \bibinfo {pages}
  {L27} (\bibinfo {year} {2016})},\ \bibinfo {note} {publisher: American
  Astronomical Society}\BibitemShut {NoStop}%
\bibitem [{\citenamefont {Andrés}\ and\ \citenamefont
  {Sahraoui}(2017)}]{andres_alternative_2017}%
  \BibitemOpen
  \bibfield  {author} {\bibinfo {author} {\bibfnamefont {N.}~\bibnamefont
  {Andrés}}\ and\ \bibinfo {author} {\bibfnamefont {F.}~\bibnamefont
  {Sahraoui}},\ }\bibfield  {title} {\bibinfo {title} {Alternative derivation
  of exact law for compressible and isothermal magnetohydrodynamics
  turbulence},\ }\href {https://doi.org/10.1103/PhysRevE.96.053205} {\bibfield
  {journal} {\bibinfo  {journal} {Phys. Rev. E}\ }\textbf {\bibinfo {volume}
  {96}},\ \bibinfo {pages} {053205} (\bibinfo {year} {2017})}\BibitemShut
  {NoStop}%
\bibitem [{\citenamefont {Andrés}\ \emph
  {et~al.}(2018{\natexlab{a}})\citenamefont {Andrés}, \citenamefont
  {Galtier},\ and\ \citenamefont {Sahraoui}}]{andres_exact_2018}%
  \BibitemOpen
  \bibfield  {author} {\bibinfo {author} {\bibfnamefont {N.}~\bibnamefont
  {Andrés}}, \bibinfo {author} {\bibfnamefont {S.}~\bibnamefont {Galtier}},\
  and\ \bibinfo {author} {\bibfnamefont {F.}~\bibnamefont {Sahraoui}},\
  }\bibfield  {title} {\bibinfo {title} {Exact law for homogeneous compressible
  {Hall} magnetohydrodynamics turbulence},\ }\href
  {https://doi.org/10.1103/PhysRevE.97.013204} {\bibfield  {journal} {\bibinfo
  {journal} {Phys. Rev. E}\ }\textbf {\bibinfo {volume} {97}},\ \bibinfo
  {pages} {013204} (\bibinfo {year} {2018}{\natexlab{a}})}\BibitemShut
  {NoStop}%
\bibitem [{\citenamefont {Andrés}\ \emph
  {et~al.}(2018{\natexlab{b}})\citenamefont {Andrés}, \citenamefont
  {Sahraoui}, \citenamefont {Galtier}, \citenamefont {Hadid}, \citenamefont
  {Dmitruk},\ and\ \citenamefont {Mininni}}]{andres_energy_2018}%
  \BibitemOpen
  \bibfield  {author} {\bibinfo {author} {\bibfnamefont {N.}~\bibnamefont
  {Andrés}}, \bibinfo {author} {\bibfnamefont {F.}~\bibnamefont {Sahraoui}},
  \bibinfo {author} {\bibfnamefont {S.}~\bibnamefont {Galtier}}, \bibinfo
  {author} {\bibfnamefont {L.~Z.}\ \bibnamefont {Hadid}}, \bibinfo {author}
  {\bibfnamefont {P.}~\bibnamefont {Dmitruk}},\ and\ \bibinfo {author}
  {\bibfnamefont {P.~D.}\ \bibnamefont {Mininni}},\ }\bibfield  {title}
  {\bibinfo {title} {Energy cascade rate in isothermal compressible
  magnetohydrodynamic turbulence},\ }\href
  {https://doi.org/10.1017/S0022377818000788} {\bibfield  {journal} {\bibinfo
  {journal} {J. Plasma Phys.}\ }\textbf {\bibinfo {volume} {84}},\ \bibinfo
  {pages} {21} (\bibinfo {year} {2018}{\natexlab{b}})}\BibitemShut {NoStop}%
\bibitem [{\citenamefont {Banerjee}\ and\ \citenamefont
  {Andrés}(2020)}]{banerjee_scale--scale_2020}%
  \BibitemOpen
  \bibfield  {author} {\bibinfo {author} {\bibfnamefont {S.}~\bibnamefont
  {Banerjee}}\ and\ \bibinfo {author} {\bibfnamefont {N.}~\bibnamefont
  {Andrés}},\ }\bibfield  {title} {\bibinfo {title} {Scale-to-scale energy
  transfer rate in compressible two-fluid plasma turbulence},\ }\href
  {https://doi.org/10.1103/PhysRevE.101.043212} {\bibfield  {journal} {\bibinfo
   {journal} {Phys. Rev. E}\ }\textbf {\bibinfo {volume} {101}},\ \bibinfo
  {pages} {043212} (\bibinfo {year} {2020})}\BibitemShut {NoStop}%
\bibitem [{\citenamefont {Ferrand}\ \emph {et~al.}(2020)\citenamefont
  {Ferrand}, \citenamefont {Galtier}, \citenamefont {Sahraoui},\ and\
  \citenamefont {Federrath}}]{ferrand_compressible_2020}%
  \BibitemOpen
  \bibfield  {author} {\bibinfo {author} {\bibfnamefont {R.}~\bibnamefont
  {Ferrand}}, \bibinfo {author} {\bibfnamefont {S.}~\bibnamefont {Galtier}},
  \bibinfo {author} {\bibfnamefont {F.}~\bibnamefont {Sahraoui}},\ and\
  \bibinfo {author} {\bibfnamefont {C.}~\bibnamefont {Federrath}},\ }\bibfield
  {title} {\bibinfo {title} {Compressible {Turbulence} in the {Interstellar}
  {Medium}: {New} {Insights} from a {High}-resolution {Supersonic} {Turbulence}
  {Simulation}},\ }\href {https://doi.org/10.3847/1538-4357/abb76e} {\bibfield
  {journal} {\bibinfo  {journal} {ApJ}\ }\textbf {\bibinfo {volume} {904}},\
  \bibinfo {pages} {160} (\bibinfo {year} {2020})},\ \bibinfo {note}
  {publisher: American Astronomical Society}\BibitemShut {NoStop}%
\bibitem [{\citenamefont {Simon}\ and\ \citenamefont
  {Sahraoui}(2021)}]{simon_general_2021}%
  \BibitemOpen
  \bibfield  {author} {\bibinfo {author} {\bibfnamefont {P.}~\bibnamefont
  {Simon}}\ and\ \bibinfo {author} {\bibfnamefont {F.}~\bibnamefont
  {Sahraoui}},\ }\bibfield  {title} {\bibinfo {title} {General {Exact} {Law} of
  {Compressible} {Isentropic} {Magnetohydrodynamic} {Flows}: {Theory} and
  {Spacecraft} {Observations} in the {Solar} {Wind}},\ }\href
  {https://doi.org/10.3847/1538-4357/ac0337} {\bibfield  {journal} {\bibinfo
  {journal} {ApJ}\ ,\ \bibinfo {pages} {9}} (\bibinfo {year}
  {2021})}\BibitemShut {NoStop}%
\bibitem [{\citenamefont {Banerjee}\ and\ \citenamefont
  {Kritsuk}(2017)}]{banerjee_exact_2017}%
  \BibitemOpen
  \bibfield  {author} {\bibinfo {author} {\bibfnamefont {S.}~\bibnamefont
  {Banerjee}}\ and\ \bibinfo {author} {\bibfnamefont {A.~G.}\ \bibnamefont
  {Kritsuk}},\ }\bibfield  {title} {\bibinfo {title} {Exact relations for
  energy transfer in self-gravitating isothermal turbulence},\ }\href
  {https://doi.org/10.1103/PhysRevE.96.053116} {\bibfield  {journal} {\bibinfo
  {journal} {Phys. Rev. E}\ }\textbf {\bibinfo {volume} {96}},\ \bibinfo
  {pages} {053116} (\bibinfo {year} {2017})},\ \bibinfo {note} {publisher:
  American Physical Society}\BibitemShut {NoStop}%
\bibitem [{\citenamefont {Banerjee}\ and\ \citenamefont
  {Kritsuk}(2018)}]{banerjee_energy_2018}%
  \BibitemOpen
  \bibfield  {author} {\bibinfo {author} {\bibfnamefont {S.}~\bibnamefont
  {Banerjee}}\ and\ \bibinfo {author} {\bibfnamefont {A.~G.}\ \bibnamefont
  {Kritsuk}},\ }\bibfield  {title} {\bibinfo {title} {Energy transfer in
  compressible magnetohydrodynamic turbulence for isothermal self-gravitating
  fluids},\ }\href {https://doi.org/10.1103/PhysRevE.97.023107} {\bibfield
  {journal} {\bibinfo  {journal} {Phys. Rev. E}\ }\textbf {\bibinfo {volume}
  {97}},\ \bibinfo {pages} {023107} (\bibinfo {year} {2018})},\ \bibinfo {note}
  {publisher: American Physical Society}\BibitemShut {NoStop}%
\bibitem [{\citenamefont {MacBride}\ \emph {et~al.}(2008)\citenamefont
  {MacBride}, \citenamefont {Smith},\ and\ \citenamefont
  {Forman}}]{macbride_turbulent_2008}%
  \BibitemOpen
  \bibfield  {author} {\bibinfo {author} {\bibfnamefont {B.~T.}\ \bibnamefont
  {MacBride}}, \bibinfo {author} {\bibfnamefont {C.~W.}\ \bibnamefont
  {Smith}},\ and\ \bibinfo {author} {\bibfnamefont {M.~A.}\ \bibnamefont
  {Forman}},\ }\bibfield  {title} {\bibinfo {title} {The {Turbulent} {Cascade}
  at 1 {AU}: {Energy} {Transfer} and the {Third}-{Order} {Scaling} for {MHD}},\
  }\href {https://doi.org/10.1086/529575} {\bibfield  {journal} {\bibinfo
  {journal} {ApJ}\ }\textbf {\bibinfo {volume} {679}},\ \bibinfo {pages} {1644}
  (\bibinfo {year} {2008})},\ \bibinfo {note} {publisher: IOP
  Publishing}\BibitemShut {NoStop}%
\bibitem [{\citenamefont {Stawarz}\ \emph {et~al.}(2009)\citenamefont
  {Stawarz}, \citenamefont {Smith}, \citenamefont {Vasquez}, \citenamefont
  {Forman},\ and\ \citenamefont {MacBride}}]{stawarz_turbulent_2009}%
  \BibitemOpen
  \bibfield  {author} {\bibinfo {author} {\bibfnamefont {J.~E.}\ \bibnamefont
  {Stawarz}}, \bibinfo {author} {\bibfnamefont {C.~W.}\ \bibnamefont {Smith}},
  \bibinfo {author} {\bibfnamefont {B.~J.}\ \bibnamefont {Vasquez}}, \bibinfo
  {author} {\bibfnamefont {M.~A.}\ \bibnamefont {Forman}},\ and\ \bibinfo
  {author} {\bibfnamefont {B.~T.}\ \bibnamefont {MacBride}},\ }\bibfield
  {title} {\bibinfo {title} {{THE} {TURBULENT} {CASCADE} {AND} {PROTON}
  {HEATING} {IN} {THE} {SOLAR} {WIND} {AT} 1 {AU}},\ }\href
  {https://doi.org/10.1088/0004-637X/697/2/1119} {\bibfield  {journal}
  {\bibinfo  {journal} {ApJ}\ }\textbf {\bibinfo {volume} {697}},\ \bibinfo
  {pages} {1119} (\bibinfo {year} {2009})},\ \bibinfo {note} {publisher:
  American Astronomical Society}\BibitemShut {NoStop}%
\bibitem [{\citenamefont {Osman}\ \emph {et~al.}(2011)\citenamefont {Osman},
  \citenamefont {Wan}, \citenamefont {Matthaeus}, \citenamefont {Weygand},\
  and\ \citenamefont {Dasso}}]{osman_anisotropic_2011}%
  \BibitemOpen
  \bibfield  {author} {\bibinfo {author} {\bibfnamefont {K.~T.}\ \bibnamefont
  {Osman}}, \bibinfo {author} {\bibfnamefont {M.}~\bibnamefont {Wan}}, \bibinfo
  {author} {\bibfnamefont {W.~H.}\ \bibnamefont {Matthaeus}}, \bibinfo {author}
  {\bibfnamefont {J.~M.}\ \bibnamefont {Weygand}},\ and\ \bibinfo {author}
  {\bibfnamefont {S.}~\bibnamefont {Dasso}},\ }\bibfield  {title} {\bibinfo
  {title} {Anisotropic {Third}-{Moment} {Estimates} of the {Energy} {Cascade}
  in {Solar} {Wind} {Turbulence} {Using} {Multispacecraft} {Data}},\ }\href
  {https://doi.org/10.1103/PhysRevLett.107.165001} {\bibfield  {journal}
  {\bibinfo  {journal} {Phys. Rev. Lett.}\ }\textbf {\bibinfo {volume} {107}},\
  \bibinfo {pages} {165001} (\bibinfo {year} {2011})}\BibitemShut {NoStop}%
\bibitem [{\citenamefont {Hadid}\ \emph {et~al.}(2017)\citenamefont {Hadid},
  \citenamefont {Sahraoui},\ and\ \citenamefont {Galtier}}]{hadid_energy_2017}%
  \BibitemOpen
  \bibfield  {author} {\bibinfo {author} {\bibfnamefont {L.~Z.}\ \bibnamefont
  {Hadid}}, \bibinfo {author} {\bibfnamefont {F.}~\bibnamefont {Sahraoui}},\
  and\ \bibinfo {author} {\bibfnamefont {S.}~\bibnamefont {Galtier}},\
  }\bibfield  {title} {\bibinfo {title} {Energy {Cascade} {Rate} {In}
  {Compressible} {Fast} {And} {Slow} {Solar} {Wind} {Turbulence}},\ }\href
  {https://doi.org/10.3847/1538-4357/aa603f} {\bibfield  {journal} {\bibinfo
  {journal} {ApJ}\ }\textbf {\bibinfo {volume} {838}},\ \bibinfo {pages} {9}
  (\bibinfo {year} {2017})},\ \bibinfo {note} {arXiv: 1612.02150}\BibitemShut
  {NoStop}%
\bibitem [{\citenamefont {Hadid}\ \emph {et~al.}(2018)\citenamefont {Hadid},
  \citenamefont {Sahraoui}, \citenamefont {Galtier},\ and\ \citenamefont
  {Huang}}]{hadid_compressible_2018}%
  \BibitemOpen
  \bibfield  {author} {\bibinfo {author} {\bibfnamefont {L.}~\bibnamefont
  {Hadid}}, \bibinfo {author} {\bibfnamefont {F.}~\bibnamefont {Sahraoui}},
  \bibinfo {author} {\bibfnamefont {S.}~\bibnamefont {Galtier}},\ and\ \bibinfo
  {author} {\bibfnamefont {S.}~\bibnamefont {Huang}},\ }\bibfield  {title}
  {\bibinfo {title} {Compressible {Magnetohydrodynamic} {Turbulence} in the
  {Earth}’s {Magnetosheath}: {Estimation} of the {Energy} {Cascade} {Rate}
  {Using} \textit{in situ} {Spacecraft} {Data}},\ }\href
  {https://doi.org/10.1103/PhysRevLett.120.055102} {\bibfield  {journal}
  {\bibinfo  {journal} {Phys. Rev. Lett.}\ }\textbf {\bibinfo {volume} {120}},\
  \bibinfo {pages} {055102} (\bibinfo {year} {2018})}\BibitemShut {NoStop}%
\bibitem [{\citenamefont {Griffel}\ and\ \citenamefont
  {Davis}(1969)}]{griffel_anisotropy_1969}%
  \BibitemOpen
  \bibfield  {author} {\bibinfo {author} {\bibfnamefont {D.}~\bibnamefont
  {Griffel}}\ and\ \bibinfo {author} {\bibfnamefont {L.}~\bibnamefont
  {Davis}},\ }\bibfield  {title} {\bibinfo {title} {The anisotropy of the solar
  wind},\ }\href {https://doi.org/10.1016/0032-0633(69)90105-6} {\bibfield
  {journal} {\bibinfo  {journal} {Planetary and Space Science}\ }\textbf
  {\bibinfo {volume} {17}},\ \bibinfo {pages} {1009} (\bibinfo {year}
  {1969})}\BibitemShut {NoStop}%
\bibitem [{\citenamefont {Gary}\ \emph {et~al.}(2001)\citenamefont {Gary},
  \citenamefont {Skoug}, \citenamefont {Steinberg},\ and\ \citenamefont
  {Smith}}]{gary_proton_2001}%
  \BibitemOpen
  \bibfield  {author} {\bibinfo {author} {\bibfnamefont {S.~P.}\ \bibnamefont
  {Gary}}, \bibinfo {author} {\bibfnamefont {R.~M.}\ \bibnamefont {Skoug}},
  \bibinfo {author} {\bibfnamefont {J.~T.}\ \bibnamefont {Steinberg}},\ and\
  \bibinfo {author} {\bibfnamefont {C.~W.}\ \bibnamefont {Smith}},\ }\bibfield
  {title} {\bibinfo {title} {Proton temperature anisotropy constraint in the
  solar wind: {ACE} observations},\ }\href
  {https://doi.org/10.1029/2001GL013165} {\bibfield  {journal} {\bibinfo
  {journal} {Geophysical Research Letters}\ }\textbf {\bibinfo {volume} {28}},\
  \bibinfo {pages} {2759} (\bibinfo {year} {2001})}\BibitemShut {NoStop}%
\bibitem [{\citenamefont {Kasper}\ \emph {et~al.}(2002)\citenamefont {Kasper},
  \citenamefont {Lazarus},\ and\ \citenamefont {Gary}}]{kasper_windswe_2002}%
  \BibitemOpen
  \bibfield  {author} {\bibinfo {author} {\bibfnamefont {J.~C.}\ \bibnamefont
  {Kasper}}, \bibinfo {author} {\bibfnamefont {A.~J.}\ \bibnamefont
  {Lazarus}},\ and\ \bibinfo {author} {\bibfnamefont {S.~P.}\ \bibnamefont
  {Gary}},\ }\bibfield  {title} {\bibinfo {title} {Wind/{SWE} observations of
  firehose constraint on solar wind proton temperature anisotropy: {FIREHOSE}
  {CONSTRAINT} {ON} {PROTON} {ANISOTROPY}},\ }\href
  {https://doi.org/10.1029/2002GL015128} {\bibfield  {journal} {\bibinfo
  {journal} {Geophys. Res. Lett.}\ }\textbf {\bibinfo {volume} {29}},\ \bibinfo
  {pages} {20} (\bibinfo {year} {2002})}\BibitemShut {NoStop}%
\bibitem [{\citenamefont {Bale}\ \emph {et~al.}(2009)\citenamefont {Bale},
  \citenamefont {Kasper}, \citenamefont {Howes}, \citenamefont {Quataert},
  \citenamefont {Salem},\ and\ \citenamefont {Sundkvist}}]{bale_magnetic_2009}%
  \BibitemOpen
  \bibfield  {author} {\bibinfo {author} {\bibfnamefont {S.~D.}\ \bibnamefont
  {Bale}}, \bibinfo {author} {\bibfnamefont {J.~C.}\ \bibnamefont {Kasper}},
  \bibinfo {author} {\bibfnamefont {G.~G.}\ \bibnamefont {Howes}}, \bibinfo
  {author} {\bibfnamefont {E.}~\bibnamefont {Quataert}}, \bibinfo {author}
  {\bibfnamefont {C.}~\bibnamefont {Salem}},\ and\ \bibinfo {author}
  {\bibfnamefont {D.}~\bibnamefont {Sundkvist}},\ }\bibfield  {title} {\bibinfo
  {title} {Magnetic {Fluctuation} {Power} {Near} {Proton} {Temperature}
  {Anisotropy} {Instability} {Thresholds} in the {Solar} {Wind}},\ }\href
  {https://doi.org/10.1103/PhysRevLett.103.211101} {\bibfield  {journal}
  {\bibinfo  {journal} {Phys. Rev. Lett.}\ }\textbf {\bibinfo {volume} {103}},\
  \bibinfo {pages} {211101} (\bibinfo {year} {2009})}\BibitemShut {NoStop}%
\bibitem [{\citenamefont {Hellinger}\ \emph {et~al.}(2006)\citenamefont
  {Hellinger}, \citenamefont {Trávníček}, \citenamefont {Kasper},\ and\
  \citenamefont {Lazarus}}]{hellinger_solar_2006}%
  \BibitemOpen
  \bibfield  {author} {\bibinfo {author} {\bibfnamefont {P.}~\bibnamefont
  {Hellinger}}, \bibinfo {author} {\bibfnamefont {P.}~\bibnamefont
  {Trávníček}}, \bibinfo {author} {\bibfnamefont {J.~C.}\ \bibnamefont
  {Kasper}},\ and\ \bibinfo {author} {\bibfnamefont {A.~J.}\ \bibnamefont
  {Lazarus}},\ }\bibfield  {title} {\bibinfo {title} {Solar wind proton
  temperature anisotropy: {Linear} theory and {WIND}/{SWE} observations},\
  }\href {https://doi.org/10.1029/2006GL025925} {\bibfield  {journal} {\bibinfo
   {journal} {Geophysical Research Letters}\ }\textbf {\bibinfo {volume}
  {33}},\ \bibinfo {pages} {L09101} (\bibinfo {year} {2006})}\BibitemShut
  {NoStop}%
\bibitem [{\citenamefont {Sahraoui}\ \emph {et~al.}(2006)\citenamefont
  {Sahraoui}, \citenamefont {Belmont}, \citenamefont {Rezeau}, \citenamefont
  {Cornilleau-Wehrlin}, \citenamefont {Pinçon},\ and\ \citenamefont
  {Balogh}}]{sahraoui_anisotropic_2006}%
  \BibitemOpen
  \bibfield  {author} {\bibinfo {author} {\bibfnamefont {F.}~\bibnamefont
  {Sahraoui}}, \bibinfo {author} {\bibfnamefont {G.}~\bibnamefont {Belmont}},
  \bibinfo {author} {\bibfnamefont {L.}~\bibnamefont {Rezeau}}, \bibinfo
  {author} {\bibfnamefont {N.}~\bibnamefont {Cornilleau-Wehrlin}}, \bibinfo
  {author} {\bibfnamefont {J.~L.}\ \bibnamefont {Pinçon}},\ and\ \bibinfo
  {author} {\bibfnamefont {A.}~\bibnamefont {Balogh}},\ }\bibfield  {title}
  {\bibinfo {title} {Anisotropic {Turbulent} {Spectra} in the {Terrestrial}
  {Magnetosheath} as {Seen} by the {Cluster} {Spacecraft}},\ }\href
  {https://doi.org/10.1103/PhysRevLett.96.075002} {\bibfield  {journal}
  {\bibinfo  {journal} {Phys. Rev. Lett.}\ }\textbf {\bibinfo {volume} {96}},\
  \bibinfo {pages} {075002} (\bibinfo {year} {2006})}\BibitemShut {NoStop}%
\bibitem [{\citenamefont {Chew}\ \emph {et~al.}(1956)\citenamefont {Chew},
  \citenamefont {Goldberger},\ and\ \citenamefont {Low}}]{chew_boltzmann_1956}%
  \BibitemOpen
  \bibfield  {author} {\bibinfo {author} {\bibfnamefont {G.~F.}\ \bibnamefont
  {Chew}}, \bibinfo {author} {\bibfnamefont {M.}~\bibnamefont {Goldberger}},\
  and\ \bibinfo {author} {\bibfnamefont {F.~E.}\ \bibnamefont {Low}},\
  }\bibfield  {title} {\bibinfo {title} {The {Boltzmann} equation an d the
  one-fluid hydromagnetic equations in the absence of particle collisions},\
  }\href {https://doi.org/10.1098/rspa.1956.0116} {\bibfield  {journal}
  {\bibinfo  {journal} {Proc. R. Soc. Lond. A}\ }\textbf {\bibinfo {volume}
  {236}},\ \bibinfo {pages} {112} (\bibinfo {year} {1956})}\BibitemShut
  {NoStop}%
\bibitem [{\citenamefont {Passot}\ and\ \citenamefont
  {Sulem}(2007)}]{passot_collisionless_2007}%
  \BibitemOpen
  \bibfield  {author} {\bibinfo {author} {\bibfnamefont {T.}~\bibnamefont
  {Passot}}\ and\ \bibinfo {author} {\bibfnamefont {P.~L.}\ \bibnamefont
  {Sulem}},\ }\bibfield  {title} {\bibinfo {title} {Collisionless
  magnetohydrodynamics with gyrokinetic effects},\ }\href
  {https://doi.org/10.1063/1.2751601} {\bibfield  {journal} {\bibinfo
  {journal} {Physics of Plasmas}\ }\textbf {\bibinfo {volume} {14}},\ \bibinfo
  {pages} {082502} (\bibinfo {year} {2007})},\ \bibinfo {note} {publisher:
  American Institute of Physics}\BibitemShut {NoStop}%
\bibitem [{\citenamefont {Hunana}\ \emph {et~al.}(2019)\citenamefont {Hunana},
  \citenamefont {Tenerani}, \citenamefont {Zank}, \citenamefont {Khomenko},
  \citenamefont {Goldstein}, \citenamefont {Webb}, \citenamefont {Cally},
  \citenamefont {Collados}, \citenamefont {Velli},\ and\ \citenamefont
  {Adhikari}}]{hunana_introductory_2019}%
  \BibitemOpen
  \bibfield  {author} {\bibinfo {author} {\bibfnamefont {P.}~\bibnamefont
  {Hunana}}, \bibinfo {author} {\bibfnamefont {A.}~\bibnamefont {Tenerani}},
  \bibinfo {author} {\bibfnamefont {G.~P.}\ \bibnamefont {Zank}}, \bibinfo
  {author} {\bibfnamefont {E.}~\bibnamefont {Khomenko}}, \bibinfo {author}
  {\bibfnamefont {M.~L.}\ \bibnamefont {Goldstein}}, \bibinfo {author}
  {\bibfnamefont {G.~M.}\ \bibnamefont {Webb}}, \bibinfo {author}
  {\bibfnamefont {P.~S.}\ \bibnamefont {Cally}}, \bibinfo {author}
  {\bibfnamefont {M.}~\bibnamefont {Collados}}, \bibinfo {author}
  {\bibfnamefont {M.}~\bibnamefont {Velli}},\ and\ \bibinfo {author}
  {\bibfnamefont {L.}~\bibnamefont {Adhikari}},\ }\bibfield  {title} {\bibinfo
  {title} {An introductory guide to fluid models with anisotropic temperatures.
  {Part} 1. {CGL} description and collisionless fluid hierarchy},\ }\href
  {https://doi.org/10.1017/S0022377819000801} {\bibfield  {journal} {\bibinfo
  {journal} {J. Plasma Phys.}\ }\textbf {\bibinfo {volume} {85}},\ \bibinfo
  {pages} {205850602} (\bibinfo {year} {2019})}\BibitemShut {NoStop}%
\bibitem [{\citenamefont {Hau}\ and\ \citenamefont
  {Wang}(2007)}]{hau_mhd_2007}%
  \BibitemOpen
  \bibfield  {author} {\bibinfo {author} {\bibfnamefont {L.~N.}\ \bibnamefont
  {Hau}}\ and\ \bibinfo {author} {\bibfnamefont {B.~J.}\ \bibnamefont {Wang}},\
  }\bibfield  {title} {\bibinfo {title} {On {MHD} waves, fire-hose and mirror
  instabilities in anisotropic plasmas},\ }\href
  {https://doi.org/10.5194/npg-14-557-2007} {\bibfield  {journal} {\bibinfo
  {journal} {Nonlinear Processes in Geophysics}\ }\textbf {\bibinfo {volume}
  {14}},\ \bibinfo {pages} {557} (\bibinfo {year} {2007})}\BibitemShut
  {NoStop}%
\bibitem [{\citenamefont {Scime}\ \emph {et~al.}(2000)\citenamefont {Scime},
  \citenamefont {Keiter}, \citenamefont {Balkey}, \citenamefont {Boivin},
  \citenamefont {Kline}, \citenamefont {Blackburn},\ and\ \citenamefont
  {Gary}}]{scime_ion_2000}%
  \BibitemOpen
  \bibfield  {author} {\bibinfo {author} {\bibfnamefont {E.~E.}\ \bibnamefont
  {Scime}}, \bibinfo {author} {\bibfnamefont {P.~A.}\ \bibnamefont {Keiter}},
  \bibinfo {author} {\bibfnamefont {M.~M.}\ \bibnamefont {Balkey}}, \bibinfo
  {author} {\bibfnamefont {R.~F.}\ \bibnamefont {Boivin}}, \bibinfo {author}
  {\bibfnamefont {J.~L.}\ \bibnamefont {Kline}}, \bibinfo {author}
  {\bibfnamefont {M.}~\bibnamefont {Blackburn}},\ and\ \bibinfo {author}
  {\bibfnamefont {S.~P.}\ \bibnamefont {Gary}},\ }\bibfield  {title} {\bibinfo
  {title} {Ion temperature anisotropy limitation in high beta plasmas},\ }\href
  {https://doi.org/10.1063/1.874036} {\bibfield  {journal} {\bibinfo  {journal}
  {Physics of Plasmas}\ }\textbf {\bibinfo {volume} {7}},\ \bibinfo {pages}
  {2157} (\bibinfo {year} {2000})},\ \bibinfo {note} {publisher: American
  Institute of Physics}\BibitemShut {NoStop}%
\bibitem [{\citenamefont {Schekochihin}\ \emph {et~al.}(2005)\citenamefont
  {Schekochihin}, \citenamefont {Cowley}, \citenamefont {Kulsrud},
  \citenamefont {Hammett},\ and\ \citenamefont
  {Sharma}}]{schekochihin_magnetised_2005}%
  \BibitemOpen
  \bibfield  {author} {\bibinfo {author} {\bibfnamefont {A.}~\bibnamefont
  {Schekochihin}}, \bibinfo {author} {\bibfnamefont {S.}~\bibnamefont
  {Cowley}}, \bibinfo {author} {\bibfnamefont {R.}~\bibnamefont {Kulsrud}},
  \bibinfo {author} {\bibfnamefont {G.}~\bibnamefont {Hammett}},\ and\ \bibinfo
  {author} {\bibfnamefont {P.}~\bibnamefont {Sharma}},\ }\bibfield  {title}
  {\bibinfo {title} {Magnetised plasma turbulence in clusters of galaxies},\
  }in\ \href@noop {} {\emph {\bibinfo {booktitle} {The {Magnetized} {Plasma} in
  {Galaxy} {Evolution}}}},\ \bibinfo {editor} {edited by\ \bibinfo {editor}
  {\bibfnamefont {K.~T.}\ \bibnamefont {Chyzy}}, \bibinfo {editor}
  {\bibfnamefont {K.}~\bibnamefont {Otmianowska-Mazur}}, \bibinfo {editor}
  {\bibfnamefont {M.}~\bibnamefont {Soida}},\ and\ \bibinfo {editor}
  {\bibfnamefont {R.-J.}\ \bibnamefont {Dettmar}}}\ (\bibinfo {year} {2005})\
  pp.\ \bibinfo {pages} {86--92},\ \bibinfo {note} {\_eprint:
  astro-ph/0411781}\BibitemShut {NoStop}%
\bibitem [{\citenamefont {Sharma}\ \emph {et~al.}(2006)\citenamefont {Sharma},
  \citenamefont {Hammett},\ and\ \citenamefont
  {Quataert}}]{sharma_shearing_2006}%
  \BibitemOpen
  \bibfield  {author} {\bibinfo {author} {\bibfnamefont {P.}~\bibnamefont
  {Sharma}}, \bibinfo {author} {\bibfnamefont {G.~W.}\ \bibnamefont
  {Hammett}},\ and\ \bibinfo {author} {\bibfnamefont {E.}~\bibnamefont
  {Quataert}},\ }\bibfield  {title} {\bibinfo {title} {Shearing box simulations
  of the {MRI} in a collisionless plasma},\ }\href@noop {} {\bibfield
  {journal} {\bibinfo  {journal} {ApJ}\ }\textbf {\bibinfo {volume} {637}},\
  \bibinfo {pages} {16} (\bibinfo {year} {2006})}\BibitemShut {NoStop}%
\bibitem [{\citenamefont {Sahraoui}\ \emph {et~al.}(2004)\citenamefont
  {Sahraoui}, \citenamefont {Belmont}, \citenamefont {Pinçon}, \citenamefont
  {Rezeau}, \citenamefont {Balogh}, \citenamefont {Robert},\ and\ \citenamefont
  {Cornilleau-Wehrlin}}]{sahraoui_magnetic_2004}%
  \BibitemOpen
  \bibfield  {author} {\bibinfo {author} {\bibfnamefont {F.}~\bibnamefont
  {Sahraoui}}, \bibinfo {author} {\bibfnamefont {G.}~\bibnamefont {Belmont}},
  \bibinfo {author} {\bibfnamefont {J.~L.}\ \bibnamefont {Pinçon}}, \bibinfo
  {author} {\bibfnamefont {L.}~\bibnamefont {Rezeau}}, \bibinfo {author}
  {\bibfnamefont {A.}~\bibnamefont {Balogh}}, \bibinfo {author} {\bibfnamefont
  {P.}~\bibnamefont {Robert}},\ and\ \bibinfo {author} {\bibfnamefont
  {N.}~\bibnamefont {Cornilleau-Wehrlin}},\ }\bibfield  {title} {\bibinfo
  {title} {Magnetic turbulent spectra in the magnetosheath: new insights},\
  }\href {https://doi.org/10.5194/angeo-22-2283-2004} {\bibfield  {journal}
  {\bibinfo  {journal} {Ann. Geophys.}\ }\textbf {\bibinfo {volume} {22}},\
  \bibinfo {pages} {2283} (\bibinfo {year} {2004})}\BibitemShut {NoStop}%
\bibitem [{\citenamefont {Kunz}\ \emph {et~al.}(2014)\citenamefont {Kunz},
  \citenamefont {Schekochihin},\ and\ \citenamefont
  {Stone}}]{kunz_firehose_2014}%
  \BibitemOpen
  \bibfield  {author} {\bibinfo {author} {\bibfnamefont {M.~W.}\ \bibnamefont
  {Kunz}}, \bibinfo {author} {\bibfnamefont {A.~A.}\ \bibnamefont
  {Schekochihin}},\ and\ \bibinfo {author} {\bibfnamefont {J.~M.}\ \bibnamefont
  {Stone}},\ }\bibfield  {title} {\bibinfo {title} {Firehose and {Mirror}
  {Instabilities} in a {Collisionless} {Shearing} {Plasma}},\ }\href
  {https://doi.org/10.1103/PhysRevLett.112.205003} {\bibfield  {journal}
  {\bibinfo  {journal} {Phys. Rev. Lett.}\ }\textbf {\bibinfo {volume} {112}},\
  \bibinfo {pages} {205003} (\bibinfo {year} {2014})},\ \bibinfo {note}
  {publisher: American Physical Society}\BibitemShut {NoStop}%
\bibitem [{\citenamefont {Osman}\ \emph {et~al.}(2013)\citenamefont {Osman},
  \citenamefont {Matthaeus}, \citenamefont {Kiyani}, \citenamefont {Hnat},\
  and\ \citenamefont {Chapman}}]{osman_proton_2013}%
  \BibitemOpen
  \bibfield  {author} {\bibinfo {author} {\bibfnamefont {K.}~\bibnamefont
  {Osman}}, \bibinfo {author} {\bibfnamefont {W.}~\bibnamefont {Matthaeus}},
  \bibinfo {author} {\bibfnamefont {K.}~\bibnamefont {Kiyani}}, \bibinfo
  {author} {\bibfnamefont {B.}~\bibnamefont {Hnat}},\ and\ \bibinfo {author}
  {\bibfnamefont {S.}~\bibnamefont {Chapman}},\ }\bibfield  {title} {\bibinfo
  {title} {Proton {Kinetic} {Effects} and {Turbulent} {Energy} {Cascade} {Rate}
  in the {Solar} {Wind}},\ }\bibfield  {journal} {\bibinfo  {journal} {Physical
  Review Letters}\ }\textbf {\bibinfo {volume} {111}},\ \href
  {https://doi.org/10.1103/PhysRevLett.111.201101}
  {10.1103/PhysRevLett.111.201101} (\bibinfo {year} {2013})\BibitemShut
  {NoStop}%
\bibitem [{\citenamefont {Hazeltine}\ \emph {et~al.}(2013)\citenamefont
  {Hazeltine}, \citenamefont {Mahajan},\ and\ \citenamefont
  {Morrison}}]{hazeltine_local_2013}%
  \BibitemOpen
  \bibfield  {author} {\bibinfo {author} {\bibfnamefont {R.~D.}\ \bibnamefont
  {Hazeltine}}, \bibinfo {author} {\bibfnamefont {S.~M.}\ \bibnamefont
  {Mahajan}},\ and\ \bibinfo {author} {\bibfnamefont {P.~J.}\ \bibnamefont
  {Morrison}},\ }\bibfield  {title} {\bibinfo {title} {Local thermodynamics of
  a magnetized, anisotropic plasma},\ }\href
  {https://doi.org/10.1063/1.4793735} {\bibfield  {journal} {\bibinfo
  {journal} {Physics of Plasmas}\ }\textbf {\bibinfo {volume} {20}},\ \bibinfo
  {pages} {022506} (\bibinfo {year} {2013})}\BibitemShut {NoStop}%
\bibitem [{\citenamefont {Kolmogorov}(1941)}]{kolmogorov_dissipation_1941}%
  \BibitemOpen
  \bibfield  {author} {\bibinfo {author} {\bibfnamefont {A.}~\bibnamefont
  {Kolmogorov}},\ }\bibfield  {title} {\bibinfo {title} {Dissipation of energy
  in locally isotropic turbulence},\ }\href@noop {} {\bibfield  {journal}
  {\bibinfo  {journal} {Dokl. Akad. Nauk SSSR}\ }\textbf {\bibinfo {volume}
  {32}},\ \bibinfo {pages} {16} (\bibinfo {year} {1941})}\BibitemShut {NoStop}%
\bibitem [{\citenamefont {Schekochihin}\ \emph {et~al.}(2009)\citenamefont
  {Schekochihin}, \citenamefont {Cowley}, \citenamefont {Dorland},
  \citenamefont {Hammett}, \citenamefont {Howes}, \citenamefont {Quataert},\
  and\ \citenamefont {Tatsuno}}]{schekochihin_astrophysical_2009}%
  \BibitemOpen
  \bibfield  {author} {\bibinfo {author} {\bibfnamefont {A.~A.}\ \bibnamefont
  {Schekochihin}}, \bibinfo {author} {\bibfnamefont {S.~C.}\ \bibnamefont
  {Cowley}}, \bibinfo {author} {\bibfnamefont {W.}~\bibnamefont {Dorland}},
  \bibinfo {author} {\bibfnamefont {G.~W.}\ \bibnamefont {Hammett}}, \bibinfo
  {author} {\bibfnamefont {G.~G.}\ \bibnamefont {Howes}}, \bibinfo {author}
  {\bibfnamefont {E.}~\bibnamefont {Quataert}},\ and\ \bibinfo {author}
  {\bibfnamefont {T.}~\bibnamefont {Tatsuno}},\ }\bibfield  {title} {\bibinfo
  {title} {Astrophysical gyrokinetics: kinetic and fluid turbulent cascades in
  magnetized weakly collisional plasma},\ }\href
  {https://doi.org/10.1088/0067-0049/182/1/310} {\bibfield  {journal} {\bibinfo
   {journal} {ApJS}\ }\textbf {\bibinfo {volume} {182}},\ \bibinfo {pages}
  {310} (\bibinfo {year} {2009})}\BibitemShut {NoStop}%
\bibitem [{\citenamefont {Sahraoui}\ \emph {et~al.}(2007)\citenamefont
  {Sahraoui}, \citenamefont {Galtier},\ and\ \citenamefont
  {Belmont}}]{sahraoui_waves_2007}%
  \BibitemOpen
  \bibfield  {author} {\bibinfo {author} {\bibfnamefont {F.}~\bibnamefont
  {Sahraoui}}, \bibinfo {author} {\bibfnamefont {S.}~\bibnamefont {Galtier}},\
  and\ \bibinfo {author} {\bibfnamefont {G.}~\bibnamefont {Belmont}},\
  }\bibfield  {title} {\bibinfo {title} {On waves in incompressible {Hall}
  magnetohydrodynamics},\ }\href {https://doi.org/10.1017/S0022377806006180}
  {\bibfield  {journal} {\bibinfo  {journal} {J. Plasma Phys.}\ }\textbf
  {\bibinfo {volume} {73}},\ \bibinfo {pages} {723} (\bibinfo {year}
  {2007})}\BibitemShut {NoStop}%
\end{thebibliography}
%

\end{document}